\documentstyle[sprocl,twoside,amsmath,amssymb]{article}

\def\beq{\begin{eqnarray}}   \def\eeq{\end{eqnarray}}
\newcommand{\nhalf}{${\cal N}\!\!=\!1/2\,$}
\newcommand{\ntwo}{${\cal N}\!\!=\!2\;$}
\newcommand{\none}{${\cal N}\!\!=\!1\,$}
\newcommand{\nfour}{${\cal N}\!\!=\!4\,$}

\def\citebk#1{\hspace{0.9mm}\raisebox{-1.85mm}[0mm][0mm]
  {\Large\cite{#1}}\hspace{-0.1mm}}

\def\citebkcap#1{\hspace{0.8mm}\raisebox{-1.5mm}[0mm][0mm]
  {\large\cite{#1}}\hspace{-0.2mm}}

\begin{document}
\sloppy

\begin{flushright}
TPI-MINN-59/01  \\
UMN-TH-2040/01\\
hep-th/0011027 \\
\end{flushright}

\vspace*{1cm}

\title{SINGLE STATE SUPERMULTIPLET IN 1+1 DIMENSIONS\,$^\dagger$}
\footnotetext{%
$^\dagger$ This contribution to the Michael Marinov Memorial Volume ``Multiple 
facets of quantization and supersymmetry'' (eds.\ M.~Olshanetsky and A.~Vainshtein, to be publish by World Scientific) presents a significant extension of 
the first version, hep-th/0011027 v.\,1. }

\author{A.~LOSEV}

\address{Institute of Theoretical and Experimental Physics,\\
B. Cheremushkinskaya 25, Moscow
117259, Russia}

\author{M.~SHIFMAN and A.~VAINSHTEIN}
\address{Theoretical Physics Institute, University of Minnesota, \\
116 Church St SE, Minneapolis, MN 55455, USA}
\maketitle

\vspace*{0.4cm}

\abstracts{
We consider multiplet shortening for BPS solitons in ${\cal N}\!\!=\!1\,$
two-dimensional models. Examples of the single-state multiplets
were established previously in ${\cal N}\!\!=\!1\,$ Landau-Ginzburg models. 
The shortening comes at a price of loosing the fermion parity $(-1)^F$ 
due to boundary effects. This implies the disappearance of the boson-fermion 
classification resulting in abnormal statistics. We discuss an appropriate 
index that counts such short multiplets. \\
\rule{4mm}{0mm} A broad class of hybrid models which extend the Landau-Ginzburg models 
to include a nonflat metric on the target space is considered. Our index
turns out to be related to the index of the Dirac operator on the soliton 
reduced moduli space (the moduli space is reduced by factoring out the
translational modulus). The index vanishes in most cases implying the absence of shortening. In particular, it
vanishes when there are only two critical points on the compact target space and the reduced moduli  space has
nonvanishing dimension. \\
\rule{4mm}{0mm} We also generalize the anomaly in the central charge to 
take into account 
the target space metric.
}

\newpage

\tableofcontents

\newpage

\setcounter{footnote}{0}

\section{Introduction}
\label{sec:intro}

The minimal supersymmetry, \none,  in two-dimensional field 
theories has 
two supercharges $Q_\alpha $  ($\alpha =1,2$) which form a Majorana spinor in
1+1 dimensions.\footnote{%
~By \none we mean what is often denoted in the literature  as  $(1,1)$ 
SUSY, with two supercharges.
} 
The centrally extended superalgebra
contains  a central charge ${\cal Z}$ which has a topological meaning.\cite{WO} 
Solitons can be defined as states with nonvanishing ${\cal Z}$.  Critical, or BPS
(Bogomol'nyi-Prasad-Sommerfield) saturated solitons are such that  their mass  
 $M$ and the corresponding central charge ${\cal Z}$ are rigidly related,
\begin{equation}
M-|{\cal Z}|=0\,.
\label{MZ}
\end{equation}
The BPS saturated
solitons preserve 1/2 of the original SUSY: one of two supercharges 
 annihilates the soliton state, while the remaining supergenerator acts nontrivially.
The irreducible representation of the superalgebra in this situation is
one-dimensional, i.e., the supermultiplet consists of a single
state.\cite{1}$^-$\cite{LSV} The shortening of the supermultiplet protects the relation (\ref{MZ})
against  small variations of 
parameters  and quantum corrections. 

If such supershort multiplet does exist (and is not accompanied by another short
multiplet), the fermion-boson classification, i.e. $(-1)^F$, is lost. 
In this paper we address this particular issue: whether or not such single-state
supermultiplets are dynamically realized. We consider
various models in  the weak coupling regime using
the quasiclassical approach. In this approach the signature of the problem is an odd
number of the fermion zero modes. In the simplest case we deal just  with one
fermion zero mode (one fermion modulus) produced by the action of the remaining
supercharge.

Let us elucidate in somewhat more detail the algebraic aspect.
The centrally extended \none superalgebra has the form
\beq
\{ Q_\alpha\,,  {\bar Q}_\beta \}=2\left(\gamma^\mu 
P_\mu +\gamma^5 {\cal Z}\right)_{\alpha\beta}\, ,
\qquad 
\left[Q_\alpha, P_\mu\right]=0\,,
\label{alg}
\eeq
where ${\bar Q}_\beta=Q_\alpha (\gamma^0)_{\alpha\beta}$ and
\begin{equation}
\gamma^0=\sigma_2\,,\qquad \gamma^1=i\sigma_3\,,\qquad
\gamma^5=i\gamma^0\gamma^1=-i\sigma_1
\label{gamma}
\end{equation}
 are purely imaginary two-by-two
matrices.

The sign of the central charge ${\cal Z}$ differentiates between solitons, ${\cal Z}>0$,
and antisolitons, ${\cal Z}<0$. In the soliton rest
frame, where $P_\mu=\left\{ {\cal Z}, 0\right\}$, it is $Q_2$ that annihilates the
soliton,
$Q_2|{\rm sol}\rangle=0$. Correspondingly, in the soliton sector the algebra reduces
to 
\begin{equation}
\left(Q_1\right)^2=2\,{\cal Z}\,,\qquad \left(Q_2\right)^2=0\,,
\end{equation}
 so there is only one supercharge $Q_1$
which  is realized nontrivially. In the semiclassical soliton construction it is just this
supercharge that generates  the fermion zero mode. The irreducible representation
of superalgebra consists of a single state 
(the representation is one-dimensional),  and
the action of the supercharge reduces to multiplication  by a number,
\begin{equation}
Q_1 \,|\,{\rm sol}\rangle=\pm \,\sqrt{2 {\cal Z}} \, |\,{\rm sol}\rangle\,.
\end{equation}

The choice of a particular sign in the irreducible representation breaks $Z_2$
symmetry which reverses the sign of the fermionic operators, 
$Q_\alpha\to -Q_\alpha$. 
 Such unusual representation implies that decomposition into bosons and
fermions (the standard
$Z_2$ grading) is broken. There are 
surprising manifestations of this phenomenon. 
For instance, in analyzing statistics
of such solitons one finds\,\cite{Tsvelik}$^-$\cite{FS} that an
effective multiplicity (defined through the entropy per soliton in the ideal
soliton gas)  is $\sqrt{2}$ instead of 1. 

Note that the problem is not specific for
supersymmetric models, it appears  always when the total number
of the fermion zero modes is odd.  In a more general context, the problem exists
even in quantum mechanics, when the Clifford algebra has odd number of
generators. In fact,  an example of such algebra is  known since long: it is  the
Grassmannian description of  nonrelativistic spin 1/2 discovered by Berezin  and
Marinov.\cite{BM}  These authors introduced three Grassmann variables, $\xi_k$
($k=1,2,3$)  which were quantized by anticommutators,
\begin{equation}
\{\xi_k\,,\xi_l\}=\delta_{kl}\,.
\label{spin}
\end{equation}

The  irreducible representations of the above algebra are two-dimensional. 
For instance, one can choose $\xi_k=\sigma_k/\sqrt{2}$.  There exists a unitary
nonequivalent choice, $\xi_k=-\sigma_k/\sqrt{2}$. The Hamiltonian $H$ depends
on $\xi$ only through the spin operators, $S_i=
-(i/2)\,\epsilon_{ikl}\,\xi_k\,\xi_l$, which are bilinear in $\xi_k$. Therefore,
the change of the sign of $\xi_k$ looks as a  classical $Z_2$ symmetry of
the problem. 

 At the quantum level, however, this symmetry is not realized:
there is no operator $G$ such that
\begin{equation}
G^2=1\,,\qquad G H G= H\,,\qquad G \xi_k G= -\xi_k\,.
\end{equation}
Alternatively, the breaking of $Z_2$ could be seen from the following. 
The operator $i\xi_1 \xi_2 \xi_3$ commutes with all $\xi_i$ and its square 
is equal to 1.  Therefore, it can be realized as $\pm 1$. The choice of a particular
sign breaks $Z_2$, the symmetry that interchanges the signs. The absence of
$Z_2$ implies the breakdown of the fermion-boson classification: the operator
trilinear in fermions is a number.

Returning to supersymmetric  field theories in
 two dimensions, let us note that the issue of the
multiplet shortening and the related loss of the fermion parity, $(-1)^F$, has a long
and sometimes confusing history.  Probably, the first encounter with the problem
occurred in the context of integrable two-dimensional models, in particular, in the
supersymmetric sine-Gordon model,\cite{Tsvelik} and in the tricritical Ising model,
field-theoretical limit of which is  the Landau-Ginzburg model with a polynomial
superpotential.\cite{Zam} Even earlier the question was discussed\,\cite{WittGN} in
the framework of the Gross-Neveu model which at $N=3$ is equivalent to the
supersymmetric sine-Gordon.

A certain clash was apparent in these considerations: on the one hand, the
existence of  $(-1)^F$ was taken for granted,\footnote{%
~In a recent private discussion E.~Witten mentioned that he had changed his opinion
as to the existence  of $(-1)^F$ in the soliton sector.
}
 implying the absence of the
single-state supermultiplets. On the other hand, as was already mentioned,  an
abnormal multiplicity factor
$\sqrt{2}$ was discovered through an entropy calculation. This unusual factor was
recently discussed anew by Witten\,\cite{WitEd} and Fendley and Saleur.\cite{FS}

The approach used in the above works was based on the exact $S$-matrices and
the thermodynamic Bethe {\em ansatz}. Close to this is the approach based on
massive deformations of conformal theories.\cite{Ahn} Another line of
development is based on the quasiclassical analysis at weak coupling. The 
main goal was
the calculation of quantum corrections to supersymmetric soliton masses,
that would 
generalize the textbook calculations of Ref.\,\protect\citebk{DHN} done in
nonsupersymmetric models. In recent years the interest to this issue was revived
by P.~van Nieuwenhuizen and collaborators.\cite{RN,NRS} These works were the
starting point for  investigations of the MIT  group\,\cite{Jaffe}
and ours.\cite{1,2,LSV} 

Already in Ref.\,\protect\citebk{1} it was mentioned that  the irreducible
representation of the superalgebra (\ref{alg}) consists of a single state, i.e. the
representation is one-dimensional. This issue was not elaborated 
in detail in Ref.\,\protect\citebk{1} although
the assertion was in contradiction with the previous
analysis of Ref.\,\protect\citebk{NRS} where it was assumed that the global $Z_2$
symmetry (associated with $(-1)^F$) is maintained.

The fact that 
the irreducible one-dimensional multiplet is realized  without doubling
was  explicitly demonstrated in Ref.\,\protect\citebk{2} where
the consideration  starts  from the  \ntwo extended version of the model.
 As was
shown in Ref.\,\protect\citebk{FI} in such \ntwo model the solitonic multiplet
is  shortened (i.e. it is two- rather than four-dimensional).
A soft breaking of \ntwo to \none  was then  introduced in
Ref.\,\protect\citebk{2}. When
the breaking parameter 
reaches some critical value, one of the two soliton states
disappears from the physical spectrum\,\cite{2} via the phenomenon  of
delocalization. Delocalization means that  fields are not localized near
the soliton center. The BPS state that remains localized is single.  The existence
of the supershort multiplets and the loss of
$(-1)^F$ was emphasized in the first version of the present paper;\,\cite{LSV2} 
in Ref.\,\protect\citebk{GLN} the authors came to the same conclusion. 

In our recent
publication\,\cite{LSV} the weak coupling
consideration was generalized to a wide
class of models including those with nonflat target space.
The present text combines (and extends) the first version of the paper 
in Ref.\,\protect\citebk{LSV2}
with  \mbox{Ref.\,\protect\citebk{LSV}}.

Let us now summarize our main points:\\[1mm]
(1) 
We introduce a new index which counts supershort multiplets in \none
two-dimensional theories. Let us remind  that 
the first SUSY index, ${\rm Tr}\, (-1)^F$, was introduced by Witten
twenty years ago\,\cite{WittenIndex} to count the number of supersymmetric 
vacua. About ten years ago, Cecotti, Fendley, Intriligator and Vafa
 introduced\,\cite{CFIV} another index,
 Tr$\,[F\,(-1)^F]$, counting the number of short multiplets in \ntwo
theories in two dimensions. No index counting single-state
multiplets in \none theories 
in two dimensions  was known. This is probably not
surprising,  since it was always assumed that $(-1)^F$ does exist.\\[1mm]  
(2) 
We will show that the appropriate index is $\left\{{\rm Tr}\,Q_1\right\}^2\!/2{\cal
Z}$ --- it vanishes for long multiplets and is equal to
1 for one-dimensional multiplets.
If the value of this index  does not vanish
in the given \none theory, 
short multiplets {\em do exist} with necessity.\\[1mm]
(3) 
We consider a wide class of \none hybrid models 
 which include,  along with a superpotential ${\cal W}(\phi)$,
a nonflat metric $g_{ab}(\phi)$ of the target space for the 
fields $\phi^a$. These are hybrids between the sigma models and 
the Landau-Ginzburg models.
The result depends on the soliton moduli space.
More exactly, what counts is a {\em reduced} moduli space, namely the moduli
space with the translational modulus (corresponding to the motion of the center
of inertia) factored out, together with its fermionic superpartner.  
The index we introduce, $\left\{{\rm Tr}\,Q_1\right\}^2\!/2{\cal
Z}$, turns out to be a square of the index of the Dirac operator defined on the
reduced moduli space.\\[1mm]
(4) 
We find,
with surprise, that  the existence of the BPS solitons belonging to
one-dimensional supermultiplets is  quite a  rare occasion. 
It happens only in the problems with a single modulus, translational.
The reduced moduli space is then trivial. Although the index of Dirac operator on a generic compact space may be nonvanishing, the reduced moduli spaces arising in the hybrid models have a special geometry (similar to spherical) for which the index vanishes. 
If the
reduced moduli space has dimension 1  or larger and is compact there are no
BPS solitons.\\[1mm]
(5) 
Another issue we address is the 
 generalization of the anomaly in the central
charge found previously in the Landau-Ginzburg models,\cite{1} where the
target space metric is flat, to the hybrid models with
 a nonflat metric. Our result for the anomaly in this
case is a straightforward extension of Ref.\,\protect\citebk{1} and can be
formulated as  a substitution 
\begin{equation}
{\cal W}(\phi) \longrightarrow \widetilde {\cal W}(\phi )={\cal
W}(\phi)+\frac{1}{4\pi}\,\nabla^a
\nabla_a {\cal W}(\phi)
\label{anoW}
\end{equation}
for the superpotential. The quantum anomaly is represented by 
the second term, with 
the covariant Laplacian on
the target space, $\nabla^a \nabla_a\equiv g^{ab}\nabla_a\nabla_b\,$.
The anomaly-corrected superpotential enters  into the energy-momentum tensor,
the supercharges and the central charge. In particular, the operator of the 
central charge  becomes
\begin{equation}
{\cal Z} = 
\widetilde{\cal W} (\phi (z\to \infty))-\widetilde{\cal W} (\phi (z\to 
-\infty))
\,.
\end{equation}

\vspace{0.2cm}

The paper is organized as follows.  In Sec.\,\ref{sec:algeb} we 
thoroughly discuss algebraic aspects. 
The relation between the multiplet shortening and the loss
of $(-1)^F$ is explained.  We also make a remark on short multiplets in
2+1 dimensions (Sec.\,2.5). Section \ref{sec:index} treats Tr$\,Q_1$ as an index.
In Sec.\,4 we begin a systematic consideration of various models.
We start from a generic hybrid model combining the
 Ginzburg-Landau and sigma models, with 
\none supersymmetry. The target space ${\cal T}$ is an arbitrary Riemann
manifold. Section 4.1 presents generalities. In Sec.\,4.2 we derive
the quantum anomaly in the central charge for the
hybrid model of the general form. Classification of the models is presented in 
Sec.\,4.3. 
In Sec.\,5 we treat particular examples with flat target space.
Section 6 is devoted to a model in which both,
the target space and the spatial coordinate are circles. 
Section 7 is devoted to nonflat target spaces.
It is demonstrated that the index Tr$\,Q_1$
is related to the index of the Dirac operator
on the soliton moduli space.
The main example here is $S^3$ as the target space where
we find the index Tr$\,Q_1$ to be vanishing.
Correspondingly, all supermultiplets are long, there are
no BPS solitons in this model.
Sections 8  summarizes our findings and results.
Appendix presents a proof  that  in any hybrid model the situation is similar:  the reduced moduli space is such that the index of the Dirac operator on it
(and, hence,  Tr$\,Q_1$) vanishes provided that the reduced moduli space 
 is a compact manifold of positive dimension. 

\section{Representations of superalgebra}
\label{sec:algeb}

\subsection{Automorphisms of superalgebra}
\label{auto}
Let us address the question: what extra symmetries are compatible with 
the centrally extended algebra  (\ref{alg})? It is clear that 
 the Lorentz boost can be included,
\begin{eqnarray}
&& Q_\alpha \to \left[\exp\left(-\frac{i}{2}\,\beta\,
\gamma^5\right)\right]_{\alpha\beta} Q_\beta\;,\qquad
\bar Q_\alpha \to \left[\exp\left(\frac{i}{2}\,\beta\,
\gamma^5\right)\right]_{\alpha\beta}\bar Q_\beta\;,
\nonumber\\[2mm]
&& P_\mu (\gamma^\mu )_{\alpha\beta}\to \left[\exp\left(-\frac{i}{2}\,\beta\,
\gamma^5\right)\right]_{\alpha\gamma}\,\left[\exp\left(\frac{i}{2}\,\beta\,
\gamma^5\right)\right]_{\beta\delta} P_\mu (\gamma^\mu
)_{\gamma \delta }\;, \nonumber\\[2mm]
&&{\cal Z}\to {\cal Z}\;,
\label{boost1}
\end{eqnarray}
where $\beta $ is a real parameter (in our convention $i \gamma^5$ is Hermitean).

If the supercharges $Q_\alpha$ were complex, as in \ntwo\!\!, an extra symmetry
would emerge --- in addition to the real $\beta $ one could consider
transformations with purely imaginary $\beta$. This symmetry would
express the conservation of the fermion charge $F$.  In \none where the
supercharges are real  there is no fermion charge. What survives, however, is the
fermion parity $G=(-1)^F$. The action of $G$ (given by putting $\beta =2i\,\pi $)
reduces to changing the sign for the fermion operators leaving the boson
operators intact,
\begin{equation}
G\,Q_\alpha\, G^{-1}= -Q_\alpha\,,\qquad G\,P_\mu\,  G^{-1}=P_\mu\,.
\end{equation}
The fermion parity $G$ realizes $Z_2$ symmetry associated with changing the
sign of the fermion fields. From this standpoint it seems that this symmetry is
guaranteed. In fact we will show in some examples that in the soliton sector the
very classification of states as either bosonic or fermionic is broken.
In constructing representations of superalgebra we would not necessarily assume
that  the fermion parity $(-1)^F$ is the valid symmetry but the Lorentz symmetry
is certainly assumed.

\subsection{Beginning the construction}
\label{begin}
Now let us start the construction.
The Lorentz symmetry implies that $P^\mu P_\mu$ is invariant,
 $P^\mu P_\mu=M^2$ where $M$ is the mass of the state. 
In what follows we will treat $M$ and ${\cal Z}$ as $c$-numbers
characterizing each given irreducible representation.
It is  convenient to choose
the rest frame  where
$P_\mu=(M,0)$ and the algebra (\ref{alg})
takes the form
\beq
(Q_1)^2 =M+{\cal Z}\,,\qquad (Q_2)^2 = M-{\cal Z}\,, \qquad \{ Q_1\,, Q_2\}
= 0\,.
\label{algone}
\eeq
Positive definiteness 
leads to $$M^2\geq {\cal Z}^2\,.$$
The analysis bifurcates at this point: one should consider separately the non-BPS
cases, $M^2> {\cal Z}^2$,  and the BPS case, $M^2={\cal Z}^2$.

Let us note that for solitons for which $M-{\cal Z}\ll {\cal Z}$
we are in domain of nonrelativistic description (it is implied that ${\cal Z}>0$). The
quantity  which plays the role of the Hamiltonian in the supersymmetric quantum
mechanics of solitons is then $M-{\cal Z}$, where $M$ should be understood as the
operator
$\sqrt{P^\mu P_\mu}\,$,
\begin{equation}
H_{\rm SQM}=M-{\cal Z}=(Q_2)^2\,.
\label{sqm}
\end{equation}
The relation $H_{\rm SQM}=(Q_2)^2$ is a subalgebra of the SUSY algebra which is
stationary for the BPS states.\footnote{~The construction (\ref{sqm}) is called
\nhalf\; quantum mechanics with
${\cal N}$ counting the number of {\em pairs} of supercharges that square to
$H_{\rm SQM}$.}

In constructing $H_{\rm SQM}$ we separate dynamics of the center of mass,
passing to the rest frame.  The total spatial momentum $P_z$ is  conjugated to  
the center of mass coordinate, $Q_1$ is its superpartner.  Both, $P_z$ and $Q_1$ 
commute with $H_{\rm SQM}$. Their partnership becomes evident in 
the moduli dynamics example which will be considered in Sec.\,\ref{sec:sigma},
zero modes generated by $P_z$ and $Q_1$  have the same dependence on the
coordinate $z$.

\subsection{Non-BPS multiplets}
\label{sec:nonbps}

If $M^2\neq {\cal Z}^2$ Eq.~({\ref{algone}}) represents  the Clifford algebra with two
generators, its
irreducible representation is two-dimensional. For instance, one can choose
\begin{equation}
Q_1=\sigma_1\sqrt{M+{\cal Z}}\,,\qquad Q_2=\sigma_2\sqrt{M-{\cal Z}}\,,
\label{two-d}
\end{equation}
where $\sigma_{1,2}$ are the Pauli matrices. 

There is an obvious automorphism: one can substitute $\sigma_1$, $\sigma_2$
by rotated matrices $\tilde \sigma_1=\sigma_1\cos\alpha +\sigma_2\sin \alpha  $
and 
$\tilde \sigma_2=-\sigma_1\sin\alpha +\sigma_2\cos \alpha  $.
It means that the two-dimensional representation at hand
admits  introduction of the fermion number $F=(1-\sigma_3 )/2 $. 
 Generically, the
fermion number operator $F$  can be expressed in terms of  the supersymmetry 
generators $Q_\alpha$.  First, let us  consider a bosonic operator
\begin{equation}
S=-\frac i 2 \,\left[Q_1\,, Q_2\right]=\frac 1 2\,
\,\bar Q\, Q\,,
\label{Rop}
\end{equation}
which is an element of the enveloping algebra.
The expression after the second equality sign is not bound to the rest frame.
This operator has the following features:
\begin{eqnarray}
&& S^2=P_\mu P^\mu-{\cal Z}^2\,,\quad \{S,Q_\alpha\}=0\,,\nonumber\\[2mm]
&& \left[S\,, Q_\alpha\right]=-2
\,\left(\gamma^\mu P_\mu +\gamma^5 {\cal Z}\right)_{\alpha\beta} Q_\beta
\nonumber\\[2mm]
&& \left[S\,, P_\mu\right]=0\,,\quad \left[S\,, {\cal Z}\right]=0\,.
\end{eqnarray}

In the representation (\ref{two-d})  the operator $S$ has the form,
\begin{equation}
S=\sqrt{M^2-{\cal Z}^2}\,\sigma_3\,.
\end{equation}
This expression shows that the operator $S$ can be introduced only for 
non-BPS representations, $M^2\neq {\cal Z}^2$. Then 
$S/\sqrt{M^2-{\cal Z}^2}$ is a generator of SO(2) rotations, associated with the
fermion charge $F$,
\begin{equation}
\frac{S}{\sqrt{M^2-{\cal Z}^2}}=1- 2\,F \,,\qquad
F=\frac{1}{2}-\frac{\bar Q Q}{4\,\sqrt{M^2-{\cal Z}^2}}\,.
\label{fchar}
\end{equation}
The fact that $S$ is bilinear in $Q_\alpha$ results in $F^2=F$, i.e. the operator $F$
acts as a projection operator. Its eigenvalues are 0 and 1 and it measures  the
number of fermions  modulo two. It means that the very same operator $S$
defines also the fermion parity $G$ of state
\begin{equation}
G=(-1)^F =\frac{\bar Q Q}{2\,\sqrt{M^2-{\cal Z}^2}}\,.
\label{gpar}
\end{equation}

Let us emphasize that in the  \none models there is no local current associated
with the fermion charge. Therefore, the fermion charge we have introduced has no
local representation.   A local  current does exists in the case of
extended \ntwo ~supersymmetry. The corresponding  
fermion charge is different from $F$ defined in Eq.~(\ref{fchar}).
 It is known that the fermion charge defined by the local current is noninteger for
solitons\,\cite{FI}  (the fractional fermion charge of the soliton was discovered by
Jackiw and Rebbi\,\cite{JR}). At the same time the fermion charge (\ref{fchar})  is
always integer.  In the topologically trivial one-particle sector of
\ntwo theories both fermion charges coincide.

Note an analogy between the introduction 
of the operator $S$ above,  in
constructing representations of the superalgebra,
 and  the  introduction
of the Pauli-Lubanski spin operator
$\Gamma^\mu=\epsilon^{\mu\nu\gamma\delta}M_{\nu\gamma}P_\delta$ for
the Poincar\'e group. In this case  there is no local current too,
and the Pauli-Lubanski operator is not defined for  
massless particles, $P_\sigma P^\sigma=0$. 
Let us emphasize once more that $S$ vanishes for the BPS states.

\subsection{BPS representations}
\label{sec:bps}

Now let us consider the special case $M^2= {\cal Z}^2$. As was mentioned,
by definition, the state for which the topological charge ${\cal Z}$ is
positive will be referred to as {\em soliton}
(${\cal Z}$ negative for {\em antisoliton}). Then for the BPS soliton $M= {\cal Z}$,
and the supercharge $Q_2$ is trivial, $Q_2=0$. 
Thus,  we are left with a single supercharge
$Q_1$ realized nontrivially. The algebra reduces to a single relation
\begin{equation}
(Q_1)^2=2\,{\cal Z}\,. 
\label{q2z}
\end{equation}
The irreducible representations of this algebra are one-dimensional, there are two
such representations,
\begin{equation}
Q_1=\pm  \,\sqrt{2{\cal Z}}\,,
\label{twosigns}
\end{equation}
i.e., two types of solitons,
\begin{equation}
Q_1|\,{\rm sol_+}\,\rangle=\sqrt{2{\cal Z}}\,|\,{\rm sol_+}\,\rangle\,,\qquad
Q_1|\,{\rm sol_{\,-}}\,\rangle=-\sqrt{2{\cal Z}}\,|\,{\rm sol_{\,-}}\,\rangle\,.
\label{1dim}
\end{equation}
It is clear that these two representations are unitary nonequivalent.

The one-dimensional irreducible representation implies
multiplet shortening: the short BPS supermultiplet contains only one state while
non-BPS supermultiplets contain two. However, the possibility of such supershort
one-dimensional multiplets is usually discarded. It is for a reason: while 
 the fermion parity $(-1)^F$ is granted in any local field theory based on fermionic
and bosonic fields, it is not defined in the one-dimensional irreducible
representation. Indeed, if it were defined, it would be
$-1$ for $Q_1$, which is incompatible with any of the equations (\ref{1dim}). 
The only way to recover 
$(-1)^F$ is to have a reducible representation containing both $|\,{\rm
sol}_{+}\,\rangle$ and $|\,{\rm sol}_{\,-}\,\rangle$.  Then, 
\begin{equation}
Q_1=\sigma _3\sqrt{2{\cal Z}}\,,\qquad
(-1)^F=\sigma_1\,. 
\label{pm}
\end{equation}

Does it mean that the one-state multiplet is not a possibility in the local field
theory? It was argued in Refs.\,\protect\citebk{1,2,GLN} (and we are going to
review this again in Sec.\,\ref{box}) that solitons in certain models do realize such
supershort multiplets indeed defying $(-1)^F$. 

Thus, for the BPS representations 
 we come to two scenarios:
\begin{itemize}
\item[{\bf (i)}]
Fermion parity $(-1)^F$ is broken and the irreducible representation is
realized. The supermultiplet is short, contains only one state.
\item[{\bf (ii)}]
Fermion parity $(-1)^F$ is not broken, the representation is reducible.
The multiplet of degenerate states is not short, containing bosonic and fermionic
components.
\end{itemize}
The important point is that  only  short multiplets  of the BPS
states are protected against becoming non-BPS under small perturbations. It is
clearly not the case in the scenario {\bf (ii)}. This leads us to introduction of a new
index, $({\rm Tr}\, Q_1)^2$, which counts short multiplets with broken $(-1)^F$.
This index will be carefully defined in Sec.\,\ref{sec:index}.

The discrete $Z_2$ symmetry $(-1)^F$ discussed above is nothing but the change of 
 sign of all fermion fields, $\psi\to -\psi$. This symmetry is seemingly present in
any theory with fermions. How this symmetry can be lost in the soliton sector will
be explained later. Here we would like to mention the following.  
Although the overall sign of $Q_1$ on the irreducible representation is not
observable, the relative sign is.
For instance, there are two types of 
 reducible representations of dimension two: one is $\{+,-\}$ (see Eq.~(\ref{pm})),
and another  $\{+,+\}$ (equivalent to $\{-,-\}$). 
Our index $(\,{\rm Tr}\, Q_1)^2/2{\cal Z}$ discriminates between these two cases
--- it vanishes in the first case and equals to 4 in the second. Another example
of dimension three is the reducible representation $\{+,+,-\}$. The index is 1 in
this case, implying that the pair  $\{+,-\}$ can leave the BPS bound leaving a
single BPS state.

\subsection{Massless supermultiplets in 2+1 dimensions}
\label{sec:2+1d}

The SUSY algebra (\ref{alg}) with the central charge ${\cal Z}$ we have considered 
in 1+1 dimensions becomes identical to the superalgebra in 2+1 dimensions
(without  central extension) provided one identifies  the central charge ${\cal Z}$
with the momentum $P_2$ in the extra spatial dimension. Indeed, after this
identification the algebra (\ref{alg}) can be rewritten as
\begin{equation}
\{ Q_\alpha\,,  {\bar Q}_\beta \}=2\left(\gamma^M\right)_{\alpha\beta} 
P_M \, ,\qquad (M=0,1,2) \, ,
\end{equation}
where $\gamma^2$ coincides with $\gamma^5$ from Eq.~(\ref{gamma}).

The one-dimensional representation we have constructed for the BPS states,
$P_\mu P^\mu={\cal Z}^2$, in 1+1 dimensions, in 2+1 becomes  a representation 
$|P_1,P_2\rangle$ for the massless particle, $P_M P^M= 0$. Assume that we choose
a Lorentz frame where $P_M=(E, 0, E)$. In this frame the supercharges
are represented as
\begin{equation}
Q_1=\pm \sqrt{2E}\,,\qquad Q_2=0\,,
\end{equation}
cf.  Eq.~(\ref{twosigns}).  Again, although irreducible
representations are one-dimensional, maintaining $(-1)^F$ makes the
representation two-dimensional and reducible, see e.g. Ref.\,\protect\citebk{BdW}. 

Is it possible to break $(-1)^F$ in 2+1D similarly to solitons in 1+1?
We are aware of no dynamical example of this type. For example, in the free
massless supersymmetric theory there are two states of the $|\pm\rangle$ type,
\begin{equation}
|P_1=0,P_2=E\rangle_\pm=\int \!{\rm d}^2 x\, {\rm e}^{i{\vec P}{\vec x}} \left(
\sqrt{2E}\,\phi \pm
\psi_1\right) |0\rangle \,,
\label{state}
\end{equation}
which are irreducible representations. Here $\phi$ and $\psi_\alpha$ are free field
operators and we chose $P_1=0$. Together they form the reducible representation
for which $(-1)^F$ is well defined.

Note that models  where the breaking of $(-1)^F$ may occur, emerge for domain
wall junctions in 3+1 dimensions.\cite{junctions} In these models the junctions
effectively reduce to 1+1 dimensional objects.

\section{Tr\,{\em Q}$_{\,1}$ as index}
\label{sec:index}

In the context of supersymmetric theories ${\rm Tr}\,(-1)^F$ as an index
 was introduced by Witten. The Witten index 
counts the difference between the numbers of the bosonic and fermionic states
of zero energy, i.e. vacua which are annihilated by supercharges.
For all supermultiplets with nonzero energy this difference  vanishes.

In particular, it vanishes in the soliton sector in \ntwo two-dimensional theories.
However, the BPS solitons are annihilated by a part of supercharges and form
short multiplets.  This is
counted\,\cite{Ntwo} by the Cecotti-Vafa-Fendley-Intriligator index  ${\rm Tr}\,
F\,(-1)^F$.  The fermion number $F$
is well defined in \ntwo theories. 

In ${\cal N}=1$ theories  the fermion number $F$, and even the fermion parity
$(-1)^F$, are not defined for short multiplets.
Is there an  index in the \none soliton problems which would count the
supershort multiplets? We assert that 
$\left\{\,{\rm Tr}\,Q_1\right\}^2$ does the job. 
More exactly,
 the  definition of the index 
 is as follows
\begin{equation}
{\rm Ind}_{\,\cal Z}\,(Q_2/Q_1)=\frac{1}{2\,{\cal Z}}\left\{\lim_{\beta \to
\infty}{\rm Tr}
\left[Q_1\,\exp(-\beta\,(Q_2)^2 )\right]\right\}^2\,.
\label{Ind}
\end{equation}
The exponential factor in Eq.\,(\ref{Ind})
is introduced for the UV regularization. 
The necessity of taking the $\beta \to\infty$ limit is due to continuous 
spectrum, as explained in Ref.\,\protect\citebk{CFIV}.

This index vanishes for non-BPS multiplets for which the fermion parity $(-1)^F$ 
can be consistently defined. Equation (\ref{two-d}) demonstrates this explicitly.
For each irreducible BPS representation the index is unity,
$$
{\rm Ind}_{\,\cal Z}\,(Q_2/Q_1)\,[\,{\rm irreducible~BPS}\,] = 1 \,.
$$
If a reducible representation contains a few irreducible BPS multiplets
the index may or may not vanish depending on the numbers of the $\{+\}$ and   
$\{-\}$ irreps. In the case of  the vanishing index one can introduce
$(-1)^F$, and small deformations or quantum corrections can destroy the BPS saturation. 

Note, that our index is not additive: it is not equal to the sum of the indices 
of the irreducible representations. 
An interesting example of an \none reducible BPS representation is provided by
solitons in \ntwo models. The \ntwo BPS multiplet consists of two \none
multiplets of the opposite types, leading to the vanishing of  the index (\ref{Ind}).

The definition (\ref{Ind}) has a technical drawback --- it refers
to the soliton rest frame. It is simple to make it Lorentz invariant,
\begin{equation}
{\rm Ind}_{\,\cal Z}\,(Q_2/Q_1)=\frac{1}{2\,{\cal Z}^2}\left(\,{\rm Tr}\,
\bar Q\right)\! \not \! P \left(\,{\rm Tr}\, Q\right)
\,,
\label{IndL}
\end{equation}
where the trace refers to the Hilbert space but not to the Lorentz indices
of the supercharges $Q_\alpha$ and $\bar Q_\alpha=Q_\beta
(\gamma^0)_{\beta\alpha}$. Here we have omitted the regularizing exponent.

\section{Theories with \none supersymmetry in 1+1 dimensions}
\label{GL}

In this section we consider a generic \none  field theory  in 1+1
dimensions, presenting a realization  of SUSY with two real supercharges
$Q_\alpha$ and the central charge, see Eq.~(\ref{alg}). 
The two-dimensional space $x^\mu=(t,z)$ is flat. The time $t\in R$ while the
spatial coordinate $z$ lives either on the line $R$ (noncompact), or on the
circle $S^1$ (compact).  We will deal with
$n$  superfields 
\begin{equation}
\Phi^a=\phi^a+\bar\theta \psi^a +\frac{1}{2\,}\bar\theta \theta F^a\,,\qquad
(a=1,\ldots,n)\,.
\end{equation}
Each superfield contains a real  boson field $\phi^a$,
a two-component Majorana  spinor $\psi_\alpha^a$ ($\alpha=1,2$)
and an auxiliary field $F^a$.  The target space formed by $\phi ^a$ is an arbitrary
Riemann manifold ${\cal T}$ endowed with the metric $g_{ab}(\phi )$.
Moreover, we introduce a superpotential ${\cal W}(\phi )$. Thus,
the generic model  is a hybrid between  the
$\sigma$-model and the Landau-Ginzburg theory.

The generic form of the Lagrangian is (for a review see
Ref.\,\protect\citebk{quantmat})
\begin{eqnarray}
{\cal L}&=&\!\frac{1}{2}\,g_{ab}\left[\,\partial_\mu\phi^a \,
\partial^\mu\phi^b + \bar\psi^a \,i\gamma^\mu {\cal D}_\mu \psi^b +F^a F^b\right]
+ \frac{1}{12} \, R_{abcd}\,(\bar\psi^a \psi^c)(\bar\psi^b \psi^d)\nonumber\\[1mm]
&+&\! F^a\partial_a {\cal W} - \frac1 2 \,(\nabla_a \partial_b {\cal W})\,\bar\psi^a
\psi^b
\,,
\label{sigmaL}
\end{eqnarray}
where  $\bar \psi=\psi^T \gamma^0$, and
$\Gamma^b_{c\,d}(\phi)$ and $R_{abcd}(\phi)$ are the Christoffel symbols
and the Riemann tensor, respectively.
Furthermore,
${\cal W}(\phi)$ is the superpotential,  and
$\partial_a$ and
$\nabla_a$ denote usual and  covariant derivatives in the target space, e.g.,
\begin{equation}
{\cal W}_{,\,a}=\nabla_a {\cal W}=\partial _a{\cal W}\,;\qquad
\nabla_a {\cal W}_{,b}=\partial_a {\cal W}_{,b} -
\Gamma^c_{ab} \, {\cal W}_{, c}\;,
\label{covar}
\end{equation}
while the covariantized space-time derivative ${\cal D}_\mu$ is
\begin{equation}
{\cal D}_\mu \, \psi^b=\frac{\partial\, \psi^b}{\partial x^\mu} +\Gamma^b_{cd}\, 
\frac{\partial\, \phi^c}{\partial x^\mu}\, \psi^d\,.
\label{covd}
\end{equation}
The Lagrangian (\ref{covar}) implies that 
the auxiliary field $F^a=-g^{ab}\partial_b{\cal W}$.

\subsection{SUSY, central charge and BPS saturation}
The \none \,supersymmetry of the model  is expressed by two supercharges,
\begin{equation}
Q_\alpha=\int {\rm d} z \,S_{\alpha}^0\,,\qquad
S^\mu=g_{ab}\left(\not \!\!\,\partial \,\phi^a-i \,F^a\right)
\gamma^\mu\,\psi^b \,,
\label{supercur}
\end{equation}
where $S^\mu$ is the conserved supercurrent.
These supercharges form the \none algebra (\ref{alg}) with the metric
independent central charge 
\begin{equation}
{\cal Z}=\int  {\rm d} z\, \partial_z \phi^a \,\partial_a {\cal W}\,.
\label{cch}
\end{equation}
The central charge does not vanish for classical solitons interpolating between
different vacua of the theory. These vacua correspond to critical points of the 
superpotential ${\cal W}(\phi)$ at which $\partial_a{\cal W}=0$.  For the soliton 
interpolating between critical points  $\phi=A$  and $\phi=B$ the central charge
is equal to 
\begin{equation}
{\cal Z}_0=\Delta{\cal W}={\cal W}(A)-{\cal W}(B)\,,
\label{claz}
\end{equation}
where we assume, by convention,  that  ${\cal Z}>0$. Certainly, the inverse
interpolation (antisoliton) with negative   ${\cal Z}$ also exists. 

For the BPS
saturated soliton and antisoliton their masses  are equal to
$|{\cal Z}|$. The BPS soliton configuration $\phi_0$ satisfies the following equation:
\begin{equation}
\frac{ {\rm d} \phi^a_{0}}{{\rm d} z}= \, g^{ab}\,\partial_b \,{\cal
W}(\phi_{0})\,.
\label{AAone}
\end{equation}

\subsection{Ultraviolet aspects and quantum anomaly}
\label{sec:anomaly}

The expressions (\ref{supercur}), (\ref{cch})   and (\ref{claz}) above 
are obtained at the classical level.
If the target space manifold ${\cal T}$ is flat, i.e., we deal with the
Landau-Ginzburg model, the theory is superrenormalizable: logarithmic
divergences appear only at one loop. For generic nonflat  ${\cal T}$
the theory is nonrenormalizable: as well known, loops will generate an
infinite series of new structures in the target space metric.  However, if 
${\cal T}$ is symmetric, the number of structures is finite, and the theory
is renormalizable: divergences can be absorbed
into  a finite number of parameters. 

Already in the superrenormalizable Landau-Ginzburg models (where
$g_{ab}=\delta_{ab}$) loop corrections lead to a quantum anomaly.\cite{1} The
anomaly occurs in the energy-momentum tensor, supercurrent and the central
charge density. In fact, these anomalies form a supermultiplet. Moreover, the
impact of the anomaly is local and universally expressed through the substitution
\begin{equation}
{\cal W}\longrightarrow \widetilde{\cal W}=
{\cal W}+\frac{1}{4\pi}\,\partial_a\partial_a{\cal W}\,.
\label{anoma}
\end{equation}

It is clear, then, that a similar anomaly must occur in the hybrid models as well.
Below we present its form for the generic model. Note, first, that 
the distinction between 
the superrenormalizable Landau-Ginzburg
models and renormalizable hybrid models is irrelevant for the analysis
of the one loop anomaly. The form of the anomaly is severely limited
by the following considerations: (i) dimension and locality;
(ii) general covariance in the target space; (iii) the  flat metric limit.
The one-loop calculation presented in Ref.\,\protect\citebk{1} can be readily
extended to include the target space metric and leads to
\begin{equation}
{\cal W} \longrightarrow \widetilde{\cal W}=
{\cal W} +\frac{1}{4\pi}\,g^{ab}\,\nabla_a\nabla_b {\cal W} 
\,.
\label{hyban}
\end{equation}
This differs from the Landau-Ginzburg case only by covariantization
of the Laplacian, 
\begin{equation}
\delta^{ab}\partial_a \partial_b{\cal W} \to 
g^{ab}\,\nabla_a \nabla_b {\cal W}=g^{-1/2}\, \partial_a
\, g^{1/2}g^{ab}\partial_b\, {\cal W}\,.
\end{equation}
The corrected superpotential $\widetilde{\cal W}$ should be substituted into
the expressions for  the energy-momentum tensor, supercurrent and the central charge.

In particular,  the central charge  ${\cal Z}$ becomes
\begin{equation}
{\cal Z}=\widetilde{\cal W}(A)-\widetilde{\cal W}(B)\,.  
\end{equation}
A novel feature compared with the flat target space is the occurrence of 
a metric dependence. Let us remind that at the loop level the metric
``runs''. It is this running metric that enters the anomaly (\ref{hyban}).
As a result, the anomaly 
which was one-loop in the flat target space\,\cite{1} becomes multi-loop.

It is worth singling out  interesting cases of superpotentials which are 
eigenfunctions of the covariant Laplacian,
\begin{equation}
g^{ab}\nabla_a\nabla_b {\cal W} = c{\cal W}\,.
\label{eigenlap}
\end{equation}
 The simplest example of this type
is provided by the sine-Gordon model, ${\cal W}=mv^2\sin(\phi /v)$ (see
Sec.\,\ref{box}).  The impact of the anomaly is the following
replacement of the classical central charge
${\cal Z}_0$:
$$
{\cal Z} ={\cal Z}_0\left[1 -\frac{1}{4\pi v^2}\right]\,.
$$
The equivalent  replacement $v^2 \to v^2-(1/4\pi )$ is well-known in the
framework of the CFT treatment of the integrable models.\cite{Ahn}

Another similar example is given by the $S^3$ model of Sec.\,\ref{sec:sigma}.
In this model the superpotential satisfies Eq.~(\ref{eigenlap})  with $c=-3f$
where $f$ is the running coupling constant.\footnote{%
~At one-loop level,
$
1/f (\mu )=1/f_0-(1/\pi)
\, \ln(\mu_0/\mu)\,.
$
}
The anomaly shifts ${\cal Z}_0$,
\begin{equation}
{\cal Z} = {\cal Z}_0 \left( 1-\frac{3f}{4\pi}
\right)\,.
\label{shifz}
\end{equation}
Equation (\ref{shifz}) implies that the combination
on the right-hand side is renor\-malization-group invariant.

\subsection{Classification of models}
\label{sec:class}

The models to be considered  in this paper fall into several distinct categories
characterized by the following features:
\begin{itemize}
\item[(i)]
The geometry of the spatial coordinate $z$: compact versus noncompact (in both
cases the metric is flat for one spatial coordinate);
\item[(ii)] The geometry of the target manifold ${\cal T}$: compact versus
noncompact.
\item[(iii)] 
The metric of ${\cal T}$: flat versus nonflat.
\end{itemize}
Our examples represent almost all combinations of the classes above.
We start with the flat target space. In this case the generic classical Lagrangian
(\ref{sigmaL}) takes the form
\begin{equation}
{\cal L}=\frac{1}{2}\left\{ \partial _\mu\phi^a
\,\partial^\mu\phi^a
+i \bar \psi^a\gamma^\mu \partial_\mu \psi^a  -\frac{\partial
{\cal W}}{\partial \phi^a}
\,\frac{\partial
{\cal W}}{\partial \phi^a} - \frac{\partial^2 {\cal W}}{\partial
\phi^a \partial
\phi^b}\, \bar \psi^a
\psi^b \right\}
\,.
\label{2lag}
\end{equation}
Some of such
models are
known to be exactly integrable.\cite{Tsvelik,Ahn,FI} However,  following
Refs.\,\protect\citebk{1}, \protect\citebk{2}, we
will limit our consideration to the quasiclassical regime assuming that  the
expansion parameter is small. 

\section{Flat target manifold, noncompact space}
\label{box}

In this  section we consider  the \none Landau-Ginzburg models
(\ref{2lag}) with one or two super\-fields with the spatial coordinate $z\in R$.

\subsection{One-superfield models}
\label{sec:onesup}

Our presentation in this section follows Ref.\,\protect\citebk{1}.  Although the
superpotential
${\cal W}(\phi)$ can be arbitrary,  for classification purposes we will discuss
two representative examples: the polynomial (PM) model,
\begin{equation}
{\cal W}_{\rm PM}(\phi ) =\frac{m^2}{4\lambda}\,\phi
-\frac{\lambda}{3}\,\phi^3\,,
\label{PS}
\end{equation}
and the sine-Gordon (SG) model,
\begin{equation}
{\cal W}_{\rm SG}(\phi ) =m v^2\sin\frac{\phi }{v}\;.
\label{SSG}
\end{equation}
The target space is noncompact in the PM case and compact $S^1$  in the SG model.
The classical BPS equation 
\begin{equation}
\frac{ {\rm d} \phi_{0}}{{\rm d} z}= {\cal
W}^\prime(\phi_0)\,,
\end{equation}
has the following solutions:
\begin{eqnarray}
& \phi_{\,0} =\mbox{\large$\frac{m}{2\lambda}$}\,{\rm tanh}\,
\mbox{\large$\frac{m\,z}{2}$}\,,~~\qquad\qquad & ({\rm PM})
\nonumber\\[2mm]
& \phi_{\,0} =v\arcsin[{\rm tanh}(mz)]\,,\qquad &
({\rm SG})
\label{solit}
\end{eqnarray}
interpolating between two neighboring  vacua. 

For infrared regularization the system is placed in a large spatial box, i.e., the
boundary conditions at
$z=\pm L/2$ are imposed.  The conditions we choose are 
\begin{eqnarray}
&&\left[\partial_z\phi -{\cal W}'(\phi) \right]_{z=\pm L/2}=0\,,\qquad
\left. \psi_1\right|_{z=\pm  L/2}=0\,,
\nonumber\\[0.2cm]
&&\left. \left[\partial_z -{\cal W}''(\phi)\right]
\psi_2\right|_{z=\pm  L/2}=0 \,,
\label{boundary}
\end{eqnarray}
where $\psi_{1,2}$ denote the components of the spinor $\psi_\alpha$.
The first line is nothing but a supergeneralization of the BPS equation, 
$D_1 \Phi(t, z=\pm L/2, \theta)=0$  at the boundary. The second line
is the  consequence of the Dirac
equation of motion, if $\psi$ satisfies the Dirac equation there is
essentially no
boundary conditions for $\psi_2\,$. Therefore, it is not an independent boundary
condition in the
solution of the classical equations of motion. We will use 
these boundary conditions later for the
construction of modes in the differential operators of the second order.

The above choice is particularly convenient because it is compatible with
the residual supersymmetry in the presence of the BPS soliton.  The boundary
conditions (\ref{boundary}) are consistent with the classical solutions,
both for
the spatially constant vacuum configurations  and for the kink. In
particular, the soliton solution
$\phi_{\,0}$ of
Eq.~(\ref{solit}) satisfies 
$\partial_z\phi - {\cal W}' =0$ everywhere.  Note that the conditions
(\ref{boundary}) are not periodic.

The next step is to introduce the expansion in modes for deviations from
the soliton
solution (\ref{solit}). For the mode expansion we use 
the  second order Hermitean differential operators  $L_2$ and
$\tilde L_2$,
\begin{equation}
L_2 = P^\dagger P\,,\qquad \tilde L_2 = P P^\dagger\,,
\end{equation}
where 
\beq
P= \partial_z -\left.{\cal W}''\right|_{\phi = \phi_0(z)},\qquad
P^\dagger= -\partial_z -\left.{\cal W}''\right|_{\phi = \phi_0(z)}\,.
\eeq
The operator $L_2$ defines the modes of $\chi
\equiv \phi-\phi_0$, and those of the
fermion field $\psi_2$, while $\tilde L_2$ does this job for $\psi_1$. The
boundary conditions for $\psi_{1,2}$ are given in Eq.~(\ref{boundary}), for
$\phi-\phi_0$ they follow from the expansion  of
the first condition in Eq.~(\ref{boundary}),
\begin{equation}
\left. \left[\partial_z -{\cal W}''(\phi_0(z))\right]
\chi\right|_{z=\pm  L/2}=0 \,.
\label{chibound}
\end{equation}

It is easy to verify that there is only one zero mode $\chi_0(z)$ for
the operator
$L_2$  which has the form,
\begin{equation}
\chi_0\propto \frac{{\rm d}\phi_0}{{\rm d}z}\propto \left.{\cal
W}'\right|_{\phi=\phi_0(z)}\propto
\left\{ \begin{array}{l}
\mbox{\large $\frac{1}{\cosh^2 (mz/2)}
$}
\quad \qquad\
~\mbox{(PM)}\,.
\\[0.4cm]
\mbox{\large $\frac{1}{\cosh\,(mz)}$}
\qquad \qquad
~~~\mbox{(SG)}\,.
\end{array}\right.
\label{chi0}
\end{equation}
This is the zero mode  for the boson field $\chi$ (translational mode)
and for
 fermion $\psi_2$ (supersymmetric mode). 

The operator $\tilde L_2$ has no zero modes at all. Let us emphasize
that the
absence of the
zero modes for $\tilde L_2$ is {\em not} because the solution of $\tilde
 L_2\,
\tilde\chi =0$ is non-normalizable (we keep the size of the box  finite) but
because of the boundary conditions $\tilde \chi (z=\pm L/2)=0$.

The translational and supersymmetric zero modes discussed above imply
that the
soliton is described by two collective coordinates: its center $z_0$ and
a ``fermionic''
center $\eta$,
\begin{equation}
\phi=\phi_0(z-z_0) +{\rm nonzero~modes } \,,\quad \psi_2=\eta\,\chi_0
+{\rm nonzero~modes}\,,
\end{equation}
where $\chi_0$ is the normalized mode given by Eq.~(\ref{chi0}). The nonzero
modes are those of the operator $L_2$.  As for $\psi_1$ it is given by
the sum over
the nonzero modes of  the operator $\tilde L_2$. 

Substituting the mode expansion
in the supercharges (\ref{supercur}) we arrive at
\begin{equation}
Q_1=2\sqrt{\cal Z}\,\eta +{\rm nonzero~modes } \,,
\qquad Q_2=\sqrt{\cal Z}\,\dot z_0\,\eta +{\rm nonzero~modes } \,.
\label{qq}
\end{equation}
Now we can proceed to the quasiclassical quantization. Projecting  the
canonic 
equal-time commutation relations for the fields $\phi$ and $\psi$ on the
zero modes  we get
\begin{equation}
\left[\,p , z_0\right]= -i\,,\qquad \eta^2=\frac 1 2\,,
\end{equation}
where $p={\cal Z}\,\dot z_0$ is the canonical momentum conjugated to
$z_0$. 
It means that in quantum dynamics of the soliton moduli $z_0$ and $\eta$
the operators
$p$ and $\eta$ can be realized as
\begin{equation}
p =-i\,\frac{\rm d}{{\rm d} z_0}\,,\qquad \eta=\frac{1}{\sqrt{2}}\,.
\end{equation}
It is clear that we could have chosen $\eta=-\,1/\sqrt{2}$. This is the same
unobservable ambiguity that was discussed in Sec.\,\ref{sec:algeb}, the supercharge
$Q_1$ is linear in $\eta.$

Thus, the supercharges depend only on the canonic momentum $p$,
\begin{equation}
Q_1=\sqrt{2\cal Z}\,,
\qquad Q_2=\frac{p }{\sqrt{2{\cal Z}}}\,.
\label{qonetwo}
\end{equation}
 In the rest frame in which we perform our consideration
$\{Q_1,Q_2\}=0$, and the only value of $p$  consistent with it is $p=0$.
Thus, for the soliton $Q_1=\sqrt{2\cal Z}$, $Q_2=0$ in full agreement
with the
general construction discussed in Sec.\,\ref{sec:algeb}.

Note that the representation (\ref{qonetwo}) can be used at nonzero $p$
as well.  It
reproduces the superalgebra (\ref{alg}) in 
the nonrelativistic limit, with $p$
having the meaning of the total spatial momentum $P_1$. 

In passing from Eq.~(\ref{qq}) to  (\ref{qonetwo}) we have omitted the nonzero
modes. For each given nonzero eigenvalue there is one bosonic eigenfunction
(in the operator $L_2$), 
the same eigenfunction  in $\psi_2$ and one eigenfunction
in $\psi_1$ (of the operator $\tilde L_2$).  The quantization of the
nonzero modes
is quite standard. The corresponding additional terms in $Q_{1,2}$ can
be easily
written in term of the creation and annihilation operators. They describe
excitations of the BPS solitons. These excitations form long (two-dimensional)
multiplets. Both supercharges do not vanish and one can introduce the fermion
number (\ref{Rop}), (\ref{fchar}).

The multiplet shortening guarantees that the equality $M={\cal Z}$ is not
corrected. For the exactly solvable \none models,\cite{Tsvelik,Ahn,FI} such as 
that with the superpotential
${\cal W}=mv^2\sin(\phi/v)$, the soliton mass is known exactly. 
In Ref.\,\protect\citebk{1} it was explicitly checked
that $M$ is   equal to the matrix element of  ${\cal Z}$ (see
Eq.~(\ref{claz}) with the account for the anomaly (\ref{anoma}))
 up to two  loops.
Moreover, it was seen
that the coupling constant expansion has a finite radius of convergence (no
essential singularity at small coupling).

What lessons can one draw from the considerations of this section? 
In the case of the polynomial model the target space is noncompact,
while the sine-Gordon case can be viewed as a compact target manifold $S^1$.
In these both cases  we found one and the
same result: short (one-dimensional) soliton multiplet defying the fermion parity.
It is clear that this conclusion remains valid for a general choice of the
superpotential ${\cal W}(\phi )$ admitting classical BPS solitons.

We would like to emphasize the following point. Although we started from a
noncompact spatial coordinate technically our analysis was performed in the finite
box (with specific boundary conditions). Thus, the infrared regularization was 
guaranteed. However,  the theory is not ultraviolet finite, only
super-renormalizable. This circumstance turn out to be crucial, as we will see in
the next section where a finite model  will be considered.

\subsection{Two-superfield model}
\label{sec:2sf}

We  start from the Landau-Ginzburg model with the extended \ntwo
super\-symmetry which is ultraviolet finite theory. Then a soft breaking 
down to \none by a mass term preserves finiteness. Our presentation in this section
follows Ref.\,\protect\citebk{2}.

The Lagrangian (\ref{2lag}) with
two real superfields $\Phi^a=\{\Phi,\widetilde \Phi\}$
has \ntwo
supersymmetry if the superpotential ${\cal W}(\phi,\widetilde \phi)$ is a
harmonic function,
\begin{equation}
\Delta_\phi {\cal W}\equiv\frac{\partial^2 {\cal W}}{\partial \phi^a \partial
\phi^a}=0\qquad \mbox{for ${\cal N}=2$}
\,.
\end{equation}
 It means, in particular, 
the absence of the anomaly in the central charge -- the
superpotential  is not changed by radiative corrections.
The \ntwo supersymmetry makes the model finite, while in \none it
was superrenormalizable. 
A polynomial example of the harmonic superpotential is 
\begin{equation}
{\cal W}(\phi,\widetilde \phi) = \frac{m^2}{4\lambda} \,\phi -
\frac{\lambda}{3} \,\phi^3 +\lambda\,\phi\,\widetilde\phi^{\,2}\,.
\end{equation}

How can one introduce  breaking  of \ntwo ? To this end,
 consider  a more general case of nonharmonic ${\cal W}\,(\phi_1,\phi_2) $,
\begin{equation}
{\cal W}\,(\phi,\widetilde \phi) = \frac{m^2}{4\lambda} \,\phi
-
\frac{\lambda}{3} \,\phi^3 +\lambda\,\phi\,\widetilde\phi^{\,2}
+\frac{p\,m}{2}\,\widetilde\phi^{\,2} +\frac{q\lambda}{3} \,\widetilde\phi^{\,3}\,,
\end{equation}
where
$p$ and $q$ are dimensionless parameters.
For $ p,q\neq 0$, the extended \mbox{\ntwo} supersymmetry
is explicitly broken down
to \none. The parameter $p$ introduces soft breaking of \ntwo
which preserves finiteness of the theory and the absence of the
anomaly.\footnote{~The term $p\,m\,\phi_2^2/2$ leads to a constant in
$\Delta_\phi {\cal W}$ which shifts the superpotential by an unobservable
constant.}
 The
nonvanishing
$q$ breaks the finiteness (the theory stays superrenormalizable,
however) and
introduces the anomaly, $\Delta_\phi {\cal W}=2q\lambda \widetilde\phi$.

The classical solution for the kink is the same as in the one-field  PM
model, see the first line in Eq.~(\ref{solit}), with  second field
$\widetilde\phi$  vanishing, 
\begin{equation}
\phi_{\rm sol}= \phi_0 (z)=\frac{m}{2\lambda}\tanh \frac{mz}{2}\,,
\qquad
\widetilde\phi_{\rm sol}=0\,.
\label{N2sol}
\end{equation}
It satisfies supersymmetric boundary conditions in the finite box, 
which has the following form in terms of superfields $\Phi^a$:
\begin{equation}
D_1 \Phi^a(t, z=\pm L/2, \theta )=0\,,\qquad a=1,2\,.
\end{equation}
It is a straightforward generalization of the one-field case (\ref{boundary}).

The mode expansion is again based on operators $L_2=P^\dagger P$,
 $\tilde L_2=P P^\dagger$ where the operator 
 \begin{equation}
   P_{ab}=\delta_{ab}\, \partial_z -\partial_a \partial_b{\cal
W}(\phi=\phi_{\rm sol})
\label{pij}
 \end{equation}
now has a matrix form. The matrix is diagonal in our case,
\begin{equation}
 P_{ab}=\left(\begin{array}{cc}
\partial_z -2\lambda\phi_0(z)&0\\[2mm]
0&\partial_z +2\lambda\phi_0(z)+pm
\end{array}\right)\,.
\end{equation}
The zero modes for the fields $\phi$, $\psi$ are the same as in 
Sec.\,\ref{box}.  A new zero mode appears in the field $\widetilde\psi$,
\begin{equation}
\left(\widetilde\psi\right)_{\rm
zero~mode}=\xi\,N \,\frac{\exp\,(-p\, mz)}{\cosh^{2}\,(mz/2)}\,\left(
\begin{array}{l}
0\\[2mm]1
\end{array}\right)\,,
\label{ps2}
\end{equation}
where $\xi$ is the operator coefficient, $N$ is the normalization factor.
At $p=0$ it has the same functional form as the old fermionic
mode in $\psi$.
This is not surprising  because  of  \ntwo supersymmetry
at $p=0$.  What is crucial is that the zero mode (\ref{ps2}) is not
lifted  even at
 nonvanishing $p$. This feature is  due to the Jackiw-Rebbi theorem.\cite{JR}

Let us consider first $q=0$ and $|p|<1$; the second condition ensures localization of
the zero mode (\ref{ps2}).  One boson and two fermion zero modes mean that we
have two soliton states which are the BPS states (within our approximation). They
form a reducible multiplet which preserves the fermion parity.  The fact that the
multiplet is not short implies that its BPS nature can be lost. It could be
demonstrated, for instance, by introducing a nonvanishing $q$.

Indeed, let us show that at $q\neq 0$ the one-loop anomaly makes $Q_2\neq 0$.
The anomalous part in $Q_2$ is
\begin{eqnarray}
Q_2&\!\!\!\!=\!\!\!\!&- \frac{1}{4\pi} \int\! {\rm d}z
\left[\frac{\partial}{\partial\phi_2 }\,
\Delta_\phi {\cal W}\right]
\left(\tilde\psi_2\right)_{\rm zero~mode}\nonumber\\[2mm]
&\!\!\!\!=\!\!\!\!&-\xi\,\frac{q\lambda\,N}{2\pi}\!\! \int\! {\rm d}z
\,\frac{\exp\,(-p\, mz)}{\cosh^{2}\,(mz/2)}=-\xi\,\frac{q\lambda}{\pi
m}\sqrt{\frac{3\,m\, p\,\pi}{2\,(1-4p^2) \tan\, (p\pi)}}
\;,
\end{eqnarray}
where $\xi$ is the second fermion modulus, $\xi^2=1/2$.
Correspondingly, the shift of the soliton mass from ${\cal Z}$ is
\begin{equation}
M-{\cal Z}=Q_2^2=\frac{3q^2\lambda^3}{4\pi^2 m}\,\frac{
p\,\pi}{(1-4p^2) \tan \,(p\pi)}\,.
\label{mz}
\end{equation}
Note the  absence of singularity at $p=1/2$.

Moreover, even at $q=0$ when there is no
anomaly (and $Q_2$ remains zero at one loop) we conjecture that a nonvanishing
$Q_2$ is generated by nonperturbative effects. If it is the case, $M-{\cal Z}\propto
\exp(-c/\lambda)$.

Now let us discuss what happens when $|p|\ge 1$.  It is clear that in this interval
the zero mode (\ref{ps2}) delocalizes: depending on the sign of $p$ it runs to the
left or right wall of the box. It becomes non-normalizable in the limit $L \to
\infty$; there is no normalization problem at finite $L$, however.
If one considers the entire system which includes the box, the supermultiplet
continues to be reducible even at   $|p|\ge 1$, i.e., unprotected against leaving the
BPS bound. 

However, physically we would like to limit ourselves to experiments which are
insensitive to the boundaries in the limit of large $L$. Then we loose one state
(associated with the boundaries) as well as the fermion parity; the multiplet
becomes short and BPS saturated at $L \to \infty$.

In fact, at $|p|\gg 1$, when the mass of the second superfield is large, this field can
be viewed as an ultraviolet regulator for the one-field model of
Sec.\,\ref{sec:onesup}.  The two-field model demonstrates that the short multiplets
appear at a price of running away from the soliton to the boundaries. The states
which run away are associated with the heavy (regulator) fields.

\section{Circle in the target and coordinate spaces}
\label{sub:circle}

In the model with Lagrangian (\ref{2lag}) with {\bf one} superfield let us
assume that the field $\phi$
lives on the circle $S^1$ of 
circumference $2\pi v$. This implies that ${\cal W}'(\phi)$
is periodic, with the period $2\pi v$. Moreover, we assume that the spatial
coordinate $z$ is also compact and defined on a circle  $S^1$ of 
circumference $L$, i.e. the points
$z$ and $z+L$ are identified. 

As was shown in Ref.\,\protect\citebk{11}, the BPS saturated solitons are  possible
provided the superpotential ${\cal W}$ is a multivalued function such
that ${\cal W}'$ is single-valued. Let us take, for instance
\beq
{\cal W}(\phi) = c\phi + w(\phi )\,,\qquad {\cal W}'(\phi)=c +w'(\phi )\,,
\eeq
where $w(\phi )$ is a $2\pi v$ periodic function and $c$ is an
appropriately chosen
numerical coefficient.  
The central charge will be equal to
$2\pi v c$.
As an example one can have in mind $w=mv^2
\sin (\phi/v)$.

The BPS equation
\vspace{-2mm}
\beq
\frac{{\rm d}\phi}{{\rm d}z} = {\cal W}'(\phi )
\eeq

\vspace{-2mm}
\noindent
has an implicit solution
\beq
\int_{\phi(0)}^{\phi(z)} \frac{{\rm d}\phi}{{\cal W}'(\phi )} = z\,. 
\label{aaa}
\eeq
The function ${\cal W}'(\phi )$ must be positive everywhere on the target
space circle. We choose the value of $\phi (0)$ such that ${\cal
W}'(\phi(0) )=
{\rm Max}\{{\cal W}'\}$, it puts the center of the soliton  at
$z=0$.

 The condition of periodicity
\beq
\int_0^{2\pi v} \frac{{\rm d}\phi}{{\cal W}'(\phi )} = L\,,
\label{mmm}
\eeq
fixes the value of $c$,
assuming that Eq.~(\ref{mmm}) has a solution, which is a generic
situation. 
We  denote the solution $\phi_0 (z)$.

The mode expansion of $\phi - \phi_0$ and $\psi_{1,2}$
is performed in the eigenmodes of differential operators $L_2$ and
$\tilde L_2$,
in the same way it was done
 in the previous section. The only difference is in the
boundary conditions. Now, instead of Eq.~(\ref{boundary}), we require
periodicity.
 In noncompact space 
the operator $L_2$ had a zero mode while
$\tilde L_2$ had no  zero mode. Now, 
in the compact space,  both have zero modes,
we denote them as $\chi_0$ for $L_2$ and $\tilde\chi_0$ for $\tilde L_2$,
\begin{eqnarray}
\chi_0&\propto &\exp\left\{\int_0^z {\cal W}''(\phi_0 (z)){\rm d}z\right\}
\propto \frac{{\rm d} \phi_0}{{\rm d} z} \propto {\cal W}'(\phi_0 )\,,
\nonumber\\[0.2cm]
\tilde \chi_0&\propto &\exp\left\{-\int_0^z {\cal W}''(\phi_0 (z))
{\rm d}z\right\}
 \propto \frac{1}{ {\cal W}'(\phi_0 )}\,.
\label{zmod}
\end{eqnarray}
Note that while the zero mode  $\chi_0$ (in $\phi$ and $\psi_2$ fields)
is localized
on the kink, the mode $\tilde \chi_0$, i.e. that of $\psi_1$ is localized
off the kink.
The zero mode balance is  the same as for nonzero modes: we have one
bosonic mode and two fermionic. 
Retaining only 
the zero modes we have the following
expansion  for the bosonic and fermionic fields:
\begin{equation}
\phi(z)=\phi_0 (z-z_0)\,,\quad
\psi_1=\xi\,\tilde\chi_0\,,\quad
 \psi_2=\eta\,\chi_0\,,
\end{equation}
where $\xi$ and $\eta$ are the fermion collective coordinates.
This leads to exactly the same supercharges as in Eq.~(\ref{qq}). The difference
lies in the
quantization relations,
\begin{equation}
\left[\,p, z_0\right]= -i\,,\quad \eta^2=\xi^2=\frac 1 2\,,\quad
\{\eta,\xi\}=0\,.
\end{equation}
Due to $\{\eta,\xi\}=0$ the representation now is two-dimensional. 

In the leading  approximation above both soliton states are BPS since
$Q_2=0$. 
However, shortly
we will show that already at the one-loop level the supercharge
$Q_2$ does not vanish. Thus, the long (two-dimensional) multiplet is
formed. The states are non-BPS, their mass exceeds the central charge by a
two-loop correction. 

The easiest way to demonstrate the phenomenon is the explicit
calculation of 
$Q_2$ with account of the anomaly (\ref{anoma}),
\begin{equation}
Q_2=\xi \int {\rm d}z \left[\partial_z \phi_0 -{\cal W}'(\phi_0)-\frac{{\cal
W}'''(\phi_0)}{4\pi}\right] \tilde\chi_0(z)\,,
\end{equation}
where we substituted the classical soliton solution for $\phi$ and the zero modes for
$\psi$ in the definition (\ref{supercur}). The zero mode of $\psi_2$
drops out
from $Q_2$ at $p={\cal Z}\dot z_0=0$. 
The term ${\cal W}'''(\phi_0)/4\pi$ is due to the
anomaly.  On the classical solution the first two terms  in the square brackets
cancel each other, only the anomalous term survives. Thus, we see that 
$Q_2\neq 0$,
\begin{equation}
Q_2=-\frac{1}{4\pi}\,\xi\,
\left[\int\frac{{\rm d} \phi}{({\cal W}')^3}\right]^{-1/2}
\int{\rm d} \phi\, \frac{{\cal W}'''}{({\cal W}')^2}\,,
\end{equation}
where we used expression (\ref{zmod}) for the zero mode $\tilde \chi_0$.
It means that the excess of the soliton mass over the central charge is
\begin{equation}
M-{\cal Z}=Q_2^2=\frac{1}{32\pi^2}\,\left[\int\frac{{\rm d} \phi}{({\cal
W}')^3}\right]^{-1}
\left[\int{\rm d} \phi\, \frac{{\cal W}'''}{({\cal W}')^2}\right]^{2}\,.
\end{equation}

Note that 
taking account of the anomaly in the model of Sec.\,\ref{box} 
(in the box) does not lead to
nonvanishing $Q_2$ because of the absence of the zero mode in $\psi_2$. 
Its  effect on $Q_1$,
\begin{equation}
\Delta Q_1=\frac{1}{\sqrt{2}}\int {\rm d}z \left[\frac{{\cal
W}'''(\phi_0)}{4\pi}\right] \chi_0(z)=\frac{1}{\sqrt{{2\cal Z}}}\,
\frac{1}{4\pi}\big[{\cal
W}''(z\to\infty)-{\cal W}''(z\to-\infty)\big]\,,
\end{equation}
amounts to the shift in 
the classical value of ${\cal Z}$ (see Eq.~(\ref{claz})) 
caused by the anomaly by virtue of the substitution
(\ref{anoma}).

\section{Nonflat  target space: \mathversion{bold} Tr$\,{Q}_{\,1}$ as  index of
the Dirac operator on the reduced moduli space}
\label{sec:nonflat}
In this section we treat target spaces with nonflat metric. 
Our central point is to show that the index ${\rm Tr} \,Q_1$ introduced above
is, in fact, the index of a Dirac operator defined  on the soliton moduli space.
More exactly, the index defined in Eq.~(\ref{Ind}) coincides with the square of the
index of a Dirac operator on the reduced moduli space of solitons (see Sec.\,7.2 for 
the definition).
The latter was studied by mathematicians.
Thus, it is possible to determine in which ${\cal N}=1$ models
${\rm Ind}_{\,\cal Z}\,(Q_2/Q_1)=0$, i.e.
the multiplet shortening does {\em not} take  place (in the general 
situation).
In particular, the index vanishes provided that the reduced moduli space is not
a point, i.e., its dimension nonvanishing, and is
 compact.  This happens, for instance, in the following situation. 
In terms of the superpotential ${\cal W}$ the soliton sweeps the interval 
$[{\cal W}(A),\,{\cal W}(B)]$. If there are no other critical points in this interval
the reduced moduli space is compact.

 A representative example of nonflat target space ${\cal T}$ is sphere $S^{n+1}$
with a superpotential producing only two critical points coinciding with the
poles of the sphere. For instance, ${\cal W}=\cos \xi $, where $\xi $ is a polar
angle, does the job.
The case $n=0$ (the circle $S^1$) was considered in 
Sec.\,\ref{sec:onesup}; this is the sine-Gordon model in which we observed short
multiplets, i.e., ${\rm Tr} \,Q_1\neq 0$. However, in the
case at hand, when we deal with a single field, the target
space metric is necessarily flat.  Note, that for higher spheres, $n\ge 1$, the
metric is necessarily nonflat.

The first nonflat example is $S^2$.  In this case we deal with two fermion moduli 
from the very beginning:
hence, $(-1)^F$ is defined and all  representations are even-dimensional. The multiplet shortening
cannot occur. This is obviously true for all odd $n$. The first example with an odd
number of fermion moduli is $S^3$ (i.e. $n=2$), from which we start.

\subsection{Superpotential on $S^{3}$ target space}
\label{sec:sigma}

In this section we consider solitons in the  model with the  sphere $S^3$ as a target
space and some specific superpotential. The target space $S^3$  is symmetric 
so the theory is contains only one running coupling.
 This example is of special  interest for us
because, as we will see,  it leads to an odd number of fermionic zero modes, similar
to the one-field model of Sec.\,\ref{box}.  We will show, however, that unlike 
the one-field model of Sec.\,5.1, in the case of $S^3$  there will be no BPS solitons.

The generic form of the Lagrangian of the sigma model is given 
by Eq.~(\ref{sigmaL}).
The metric in this case is given by the following
expression for the interval,
\begin{equation}
{\rm d} s_3^2=g_{ab}\,{\rm d}\phi^a {\rm d}\phi^b=\frac{1}{\lambda}\left[{\rm
d}\xi^2 +\sin^2\xi\left({\rm d}\theta^2 +\sin^2\theta {\rm
d}\varphi^2\right)\right]\,,
\end{equation}
where $\lambda$ is the coupling
constant and we choose the  angle coordinates $\xi$, $\theta$ and $\varphi$ to
parameterize
$S^3$. 

 The superpotential is 
\begin{equation}
{\cal W}(\phi)=\frac{1}{2}\,M_0\cos \xi\,.
\end{equation}
The superpotential has maximum at $\xi=0$ and minimum at $\xi=\pi$ and no
other critical points.  The critical points are two vacua of the theory. Excitations at
these vacua form three boson-fermion supermultiplets with the mass
$\lambda M_0/2$.

In the classical approximation the model has a family of BPS solitons
interpolating between the maximum and the minimum.  The mass of the soliton is
given by the central charge (\ref{cch})
\begin{equation}
M_{\rm sol}={\cal Z}=\Delta {\cal W}=M_0\,.
\end{equation}
Its profile as a function of the coordinate $z$ is
determined by the BPS equations,
\begin{equation}
\frac{{\rm d} \phi^a_{\rm sol}}{{\rm d} z}=g^{ab}\,\partial_b \,{\cal
W}(\phi_{\rm sol})\,.
\end{equation}
 On the soliton trajectories
$\xi$ changes between 0 and $\pi$ at fixed $\theta$ and $\phi$,
\begin{eqnarray}
&&\xi_{\rm sol}(z) =2 \arctan\left(\exp\left[-\frac 1 2 \,\lambda M_0 \,
(z-z_0)\right]\,\right)\,,\nonumber\\[1mm]
&& \theta_{\rm sol}(z)=\theta_0\,,\quad \varphi_{\rm sol} (z)=\varphi_0\,,
\end{eqnarray}
so in
addition to the soliton center $z_0$ there are two  extra moduli,
$\theta_0$ and $\varphi_0$.  We denote the set of all three moduli by
$m^i=\{\theta_0,\varphi_0,z_0\}$ where $i=1,2,3$.

There is a fermionic partner $\eta^{i}$ to
each bosonic modulus and the 
 fields  in the soliton sector are
represented as 
\begin{eqnarray}
\label{fieldexp}
&&\!\!\!\!\!\phi^a(z, t)=\phi^a_{\rm sol}(z, \, m^i) +\mbox{nonzero modes}\,,
\qquad a=1,2,3\\[1mm]
&&\!\!\!\!\!\psi^a_1(z, t)=\mbox{nonzero
modes}\,,\quad\psi^a_2(z, t)=
 \eta^i \, \frac{\partial \phi^a_{\rm sol} }{\partial m^i}+\mbox{nonzero
modes}\,,\nonumber
\end{eqnarray}
where the time dependence enters through collective coordinates.
Equation (\ref{fieldexp}) implies that we are in the rest frame of the soliton.

Substituting these expressions in the Lagrangian (\ref{sigmaL}) and neglecting all
nonzero modes we obtain the Lagrangian for dynamics of moduli,
\begin{equation}
L=
\frac{1}{2}\,h_{ij}(m) \left( \dot m^i \,\dot m^j +
i\,\eta^i \,{\cal D}_t\,\eta^j\right)-M_0\,,
\label{lmod}
\end{equation}
where the induced metric $h_{ij}(m)$ refers to the  $S^2\times R $
 geometry of the moduli space,
\begin{equation}
{\rm d} s_m^2= h_{ij}(m) \,{\rm d} m^i \,{\rm d} m^j=
 \frac{4}{\lambda^2\,M_0}\left[{\rm d}\theta_0^2 +\sin^2\!
\theta_0\, {\rm d}\varphi_0^2\right] + M_0\, {\rm d}z_0^2\,.
\label{meth}
\end{equation}
The coordinate on $R$ is $m^3=z_0$  and
$m^{1,2}=\{\theta_0,
\varphi_0\}$ are the angles on $S^2$.
Moreover,  the covariant derivative ${\cal D}_t$ is defined as
\begin{equation}
{\cal D}_t\,\eta^j=\dot \eta^j +\dot m^k \,\tilde\Gamma^j_{kl} \,\eta^l\,.
\end{equation}
This is  in correspondence with the field-theoric definition (\ref{covd}), but 
the   Christoffel symbols $\tilde\Gamma^j_{kl}$ refer, of course, to the moduli 
metric $h_{ij}$. We put the tilde to differentiate from the field theory
ones.

By the same token we get also expressions for the supercharges,
\begin{eqnarray}
&& Q_1=2{\cal Z}\, \eta^1\,,\label{q1}\\[2mm]
&& Q_2= h_{ij}\, \dot m^i \,\eta^j = \dot m_j \,\eta^j\,,\label{q2}
\end{eqnarray}
in terms of  bosonic and fermionic moduli where ${\cal Z}=M_0$.

The  next step is to quantize the moduli dynamics.\footnote{~The procedure of
quantization has a rich literature. In our presentation we  follow 
the one given in Witten's lecture.\cite{WL} As far as technical  details are
concerned, we closely follow \mbox{Ref.\,\protect\citebkcap{Ali}} where they are
thoroughly discussed.}  To this end one introduces  the canonic momenta
conjugated to the coordinates and imposes commutation relations. The crucial point
is to establish the ordering of noncommuting operators.  The ordering is completely
fixed by general covariance in the target space and supersymmetry. Namely, unlike
Hamiltonian, quantization of the supercharge is uniquely defined.

 For the  bosonic coordinates $m^i$ one has
\begin{equation}
p_i =\frac{\partial L}{\partial \dot m^i}=h_{ij}\,\dot m^j +\frac{i}{2}\,\eta_j\,
\tilde\Gamma^j_{i\,l} \, \eta^l\,,\qquad [p_i\, , m^j]=-i\delta^j_i\,.
\label{comm}
\end{equation}
For the fermion coordinates $\eta^i$,
\begin{equation}
\zeta_i=\frac{\partial L}{\partial \dot \eta^i}=i\,h_{ij}\,\eta^j\,,
\qquad \{\zeta_i\, , \eta^j\}=i\,\delta^j_i\,,
\qquad \{\eta^i\, , \eta^j\}= h^{ij}\,.
\label{antico}
\end{equation}
A subtlety in this case is that the canonic momenta are function of coordinates and 
are not independent. The validity of the anticommutation relations (\ref{antico})
can be verified by substituting expressions (\ref{fieldexp}) into field-theoretic
commutators. Alternatively, one can check them  considering Green functions
in quantum mechanics (see, e.g., Ref.\,\protect\citebk{Ali}). 

We can realize the algebra of the commutation relations (\ref{comm}), 
(\ref{antico}) in the Hilbert space of two-component spinor wave functions  $\Psi_a
(m)$ with the scalar product 
\begin{equation}
\langle\Phi\, | \,\Psi \rangle =\int\! \Pi {\rm d} m^i
\, \sqrt{h(m)}\; \Phi^\dagger (m) \, \Psi(m) 
\end{equation}
in the following way
\begin{eqnarray}
&& p_i=-i\,\delta_{ab}\, {h}^{-1/4}\,\frac{\partial}{\partial m^i}\,{h}^{1/4}\,,
\qquad (i=1,2,3)\,, \quad (a,b=1,2)\,,\nonumber\\[1mm]
&& \eta^i\equiv\frac{1}{\sqrt{2}}\,\sigma^i\equiv\frac{1}{\sqrt{2}}\,e^i_A
\left(\sigma^A\right)_{ab}\,,~~~
\qquad \qquad (A=1,2,3)\,, 
\label{modsigma}
\end{eqnarray}
where $\sigma^A$ are the Pauli matrices and we introduce frames $e^i_A$,
satisfying the conditions $e^i_A e^i_B \, \delta^{AB}=h^{ij}$. A possible choice 
for $e^i_A$ is
\begin{equation}
e^i_A={\rm diag} \left\{\frac{\lambda\sqrt{{\cal Z}}}{2}
\;,
\frac{\lambda\sqrt{{\cal Z}}}{2\, \sin \theta_0}\;,  \frac{1}{\sqrt{{\cal Z}}}\right\}\,.
\end{equation}

In fact, what we need is the quantum version of the classical supercharges
presented in Eqs.~(\ref{q1}),  (\ref{q2}),
\begin{eqnarray}
&& Q_1=2{\cal Z}\, \eta^1\,,\label{qq1}\\[1mm]
&& Q_2= \frac{1}{2}\left(\eta^i \,\pi_i +\pi_i\,\eta^i\right)\,,\label{qq2}
\end{eqnarray}
where the operator of covariant momentum $\pi_i$ (a quantum version of
the velocity operator $\dot m_i=h_{ij}\dot m^j$, see Eq.~(\ref{comm})) is
defined as 
\begin{equation}
\pi_i=p_i -\frac{i}{4}\,\tilde\Gamma_{j,i\,l} \left[\eta^j,\eta^l\right]\,.
\label{pii}
\end{equation}
In fact, $\pi_i$ reduces to the covariant derivative on the spin manifold,
$\pi_i=-i\nabla_i\,$, the fermion term in Eq.~(\ref{pii}) represent the spin
connection.  In terms of $\nabla_i$ the supercharge (\ref{qq2}) can be rewritten as
\begin{equation}
Q_2= \eta^j \; (-i \nabla_j)\,\equiv \frac{1}{\sqrt{2}}\,\sigma^j\,(-i\nabla_j)\,,
\qquad j=1,2\,.
\label{dirop}
\end{equation}
This is nothing but the Dirac operator $\not \!\! D$ on the manifold.

Let us stress that the dynamics of the moduli $m^3$, $\eta^3$ along the $R$
direction is factored out, this is just a free motion of the center of mass (together 
with its fermionic partner). In particular $\pi^3\! \equiv p^3$ is conserved and we
set it to zero by choosing the rest frame. It means that the sum in
Eqs.~(\ref{qq2}), (\ref{dirop}) for $Q_2$ runs only over the $S^2$ coordinates,
$i=1,2$. We will show below that the situation is general: the moduli space always factorizes as ${\cal R}\otimes {\cal M}$.

The commutators 
\begin{eqnarray}
\left[\pi_i\,, m^j\right]=-i\delta^j_i\,, \qquad
\left[\pi_i\,, \pi_j\right]=-\frac 1 2\,\tilde R_{ij kl}\,\eta^k\,\eta^l\,,
\qquad \left[\pi_i\,, \eta^j\right]=i\,\tilde\Gamma^j_{i\,l} \,\eta^l
\end{eqnarray}
allow one to calculate the commutators of supercharges (\ref{qq1}), (\ref{qq2})
 with the coordinates $m^i$, $\eta^i$,
\begin{eqnarray}
&&\left[Q_1, m^i\right]=0\,,\qquad \left\{Q_1, \eta^1\right\}=2\,,\qquad
\left\{Q_1, \eta^{2,3}\right\}=0\,,\nonumber\\[1mm]
&&\left[Q_2, m^i\right]=-i\,\eta^i\,,\qquad 
\left\{Q_2, \eta^i\right\}=\frac{1}{2}\left(h^{ij}\,\pi_j+ \pi_j \,h^{ij}\right)\,.
\end{eqnarray}
These commutators match the classical supersymmetry transformations.

Finally, the algebra of the supercharges $Q_{1,2}$ is
\begin{eqnarray}
&& (Q_1)^2= 2\,{\cal Z}\,,\qquad \{Q_1,Q_2\}=0\label{hamilt1}\,,\\[2mm]
&& (Q_2)^2= H-{\cal Z}=\frac{1}{2}\,h^{-1/4}\,\pi_i\,h^{1/2}\, h^{ij}\,\pi _j\,h^{-1/4}
+\frac{1}{8}\,\tilde R\,.
\label{hamilt}
\end{eqnarray}
The expression (\ref{hamilt}) for $(Q_2)^2\!=(1/2)(\sigma^j\;i\nabla_j )^2$ is 
a particular case of the
famous Lichnerowicz formula: the first term can be written as
$-h^{ij}\nabla_i\nabla_j/2$, i.e., it  represents  the
invariant Laplacian  in application to spinors,  and 
 $\tilde R$ denotes the scalar curvature for the moduli metric $h_{ij}$.
In our example it is the curvature of the $S^2$ sphere,
\begin{equation}
\tilde R=\frac{\lambda^2}{2}\,{\cal Z}\,.
\end{equation}

Although our derivation was framed in terms of a concrete metric the results
(\ref{hamilt1}), (\ref{hamilt}) are perfectly general and  can be applied to 
$\sigma$ model on arbitrary manifold. The geometry of the moduli space, i.e.,
$h_{ij}$, depends on  both: the field-theoretic metric $g_{ab}$ and on the form of
the superpotential ${\cal W}$.

Let us return to our example. In this case the scalar curvature $\tilde R$
 is clearly a positive constant. This provides a positive term in the
Hamiltonian, the last term in Eq.~(\ref{hamilt}). The first term in $H-{\cal Z}$,
which  coincides (up to fermion terms) with  the invariant Laplacian, is positive
definite by itself. Thus, there can be no zero eigenvalues of
$H-{\cal Z}$. In other words, there are no states which are annihilated by the
supercharge $Q_2$. The states which were BPS saturated at the classical level cease
to be BPS at the quantum level. 

It is not difficult to determine a complete spectrum of $H-{\cal Z}$ in the $S^2$ case.
We limit ourselves to the lowest eigenvalue.
After some simple algebra we get 
\begin{equation}
\frac{4}{\tilde R} \,h^{-1/4}\,\pi_i\,h^{1/2}\, h^{ij}\,\pi _j\,h^{-1/4}=-\frac{1}{\sin
\theta}\,\frac{\partial}{\partial\theta}\,\sin \theta
\,\frac{\partial}{\partial\theta}+ \frac{1}{\sin^2 \theta}\left(i\,
\frac{\partial}{\partial\varphi} +\frac{\sigma^3}{2} \cos\theta\right)^2 .
\nonumber
\end{equation}
The $\varphi$ dependence is just $\exp(i\,m\,\varphi)$ and for the ground state
$m=0$. The $\theta$ dependence is given by the associated Legendre functions
$P^{\pm1/2}_{1/2}$, so we get the lowest eigenvalue  equal to $3\tilde
R/16$  and doubly degenerate ground state,
\begin{eqnarray}
&&\Psi_{1/2}= \sqrt{\frac{\sin\theta}{\pi}}\left(\begin{array}{cc}1\\ 0
\end{array} \right)\,,\quad
\Psi_{-1/2}=
\sqrt{\frac{\sin\theta}{\pi}}\left(\begin{array}{cc}0\nonumber\\ 1
\end{array} \right)\,,\\[1mm]
&&\left(H-{\cal Z}\right)\Psi_{\pm1/2}= \frac{3}{4}\cdot \frac{\tilde
R}{4}\,\Psi_{\pm1/2}\,.
\end{eqnarray}

The moduli dynamics we have considered is a nice example of the ${\cal
N}\!=\!1/2$ supersymmetric quantum mechanics. The possibility of a nontrivial 
${\cal N}\!=\!1/2$ construction is due to the fact that the interaction enters through
kinetic terms rather than through potential. The same moduli dynamics can be
viewed as a theory based on the Dirac operator defined on the curved moduli space.
Indeed, the realization of $ H-{\cal Z}=(Q_2)^2$ by matrix-valued differential
operators in the moduli dynamics allows to interpreted the supercharge $Q_2$
as the Dirac operator on the moduli manifold.

The  Dirac operators on manifolds were extensively studied in the mathematical
literature. In particular, the absence of zero modes of the Dirac operator on $S^n$
spheres (in our language the absence of the BPS states) follows from
the Lichnerowicz formula (\ref{hamilt}) as was mentioned above. More generally,
one can introduce an index of the Dirac operator $ \not\!\! D$ which counts the
difference of left and right chiral zero modes,
\begin{equation}
{\rm ind}\,(\not \!\! D)={\rm Tr} \left[\,\sigma^3 \exp\left(-\beta\not \!\! D^2
\right)\right]\,,
\label{index}
\end{equation}
where the matrix $\sigma^3$  anticommutes with the Dirac operator $\not \!\!
D$  defined by Eq.~(\ref{dirop}).  For $S^2$ the Dirac operator has no zero modes 
at all, so the index vanishes. 

The matrix $\sigma^3$ in the definition (\ref{index}) is a realization of
$\gamma^5$ in our $S^2$ case. Moreover,  in the moduli dynamics $\sigma^3$
is a realization of the supercharge $Q_1=\sqrt{2{\cal Z}}\,\sigma^3 \,$, therefore the
index can be rewritten  as
\begin{equation}
{\rm ind}\,(\not \!\! D)={\rm Tr} \left[\frac{Q_1}{\sqrt{2{\cal
Z}}}\,\exp\left(-\beta\,Q_2^2\right)\right]\,.
\end{equation}

\subsection{ Generic target space}

Let us pass now to the general case: ${\cal T}$ is an arbitrary Riemann manifold
endowed with a metric $g_{ab}(\phi )$ and a superpotential ${\cal W}(\phi)$. The
classical vacua are the critical points of the superpotential, $\partial_a {\cal
W}=0$. Classical BPS solitons interpolate between vacua $A$ and $B$ and satisfy
the first order differential equations (\ref{AAone}).

We start from  briefly reviewing elements of the Morse theory
(see e.g. Refs.\,\protect\citebk{novem2}, \protect\citebk{novem3}).
For every critical point $A$ the Morse index of this point $\nu(A)$
is defined as the number of the negative eigenvalues in 
the matrix of the second derivatives 
\begin{equation}
H_{ab}(\phi)=\nabla_a \partial_b {\cal W}(\phi)
\end{equation}
at $\phi =A$. At the critical points 
the covariant derivative $\nabla_a$ coincides with the regular $\partial_a$.
For solitons interpolating between two critical points, $\phi =A$ at 
$z\to -\infty$
and $\phi =B$ at $z\to \infty$
one can determine the relative Morse index $\nu_{BA}$,
\begin{equation}
\nu_{BA} = \nu (B) - \nu (A)\,.
\end{equation}
This relative Morse index counts the difference between the numbers of the zero
modes  of the operators $P$ and $P^\dagger$,
\begin{equation}
\nu_{BA}={\rm ker}\left\{P\right\}- {\rm ker}\left\{P^\dagger\right\}\,,
\end{equation}
where $P$ and $P^\dagger$ are 
\begin{equation}
P_{ab}= g_{ab} D_z - H_{ab}\,,\qquad P^\dagger_{ab}= -g_{ab} D_z - H_{ab}\,.
\end{equation}
Here $D_z$ is defined in Eq.~(\ref{covd}), and the field $\phi$ is taken to be 
$\phi_{\rm sol}(z)$. 

Note that $\nu_{BA}=0$
in the \ntwo case and its small deformations. Indeed, due to the
harmonicity of ${\cal W}$ in this case, $\Delta_\phi {\cal W}=0$,
which leads to one negative eigenvalue in the matrix of the second
derivatives in each vacua (per pair of fields related by \ntwo).

For the BPS soliton, satisfying Eq.~(\ref{AAone}), one zero mode certainly 
present
in $P$ is the translational mode. It corresponds to the soliton center 
$z_0$, one of
the coordinates in the soliton moduli space. The same zero mode of  $P$  is the
fermion zero mode --- the corresponding modulus $\eta$ is the superpartner of
$z_0$.

We will limit ourselves to the case when $ {\rm ker}\,{P^\dagger}=0$.
(Note that even if that is not the case, one can get rid of the zero modes in
$P^\dagger$ by small deformations of the superpotential). Then, the Morse 
index 
\begin{equation}
\nu_{BA}\equiv n+1 \ge 1
\end{equation}
counts  the dimension of the soliton moduli space
${M}^{n+1}$. Thus, we arrive at quantum mechanics of $n+1$ bosonic
and $n+1$ fermionic moduli on ${M}^{n+1}$.

As was mentioned above, one of $n+1$ bosonic
 moduli is $z_0$, the  coordinate of
the soliton center.
This is a cyclic coordinate conjugated to 
the generator $P_z$ of the spatial translations, $z_0 \in  {\cal R}$.
Note an ambiguity in $z_0$ --- one can add to $z_0$ an arbitrary function of 
other moduli. This ambiguity is fixed by the definition given below, see
Eq.~(\ref{mzero}).  Thus, 
the moduli space ${M}^{n+1}$ is a direct product
\begin{equation}
{M}^{n+1}={\cal R}\otimes {\cal M}^{n}
\label{RM}
\end{equation}
of ${\cal R}$ and the manifold ${\cal M}^{n}$ with coordinates  $m^1,...,m^{n}$
describing internal degrees of freedom of the soliton.  This manifold ${\cal
M}^{n}$  is what we call the {\em reduced moduli space}. 

It is instructive to elucidate the
factorization (\ref{RM}) in more detail.\cite{LSV}
We must show that the moduli space metric
$h_{ij}$,
\begin{equation}
h_{ij}(m) =\!\int\! {\rm d}z\, g_{ab}(\phi_{\rm sol})\,\,
\frac{\partial \phi_{\rm sol}^a}{\partial m^n}\,\,
\frac{\partial \phi_{\rm sol}^b}{\partial m^k}\,,\qquad i,j=0,1,2,...,n\,,
\label{Aone}
\end{equation}
where $m^0\equiv z_0$, has a block form,
i.e. $h_{0j}=0$ for $j=1,2,...,n$.
Indeed,
\begin{equation}
h_{0j}(m)\, =- \!\int\! {\rm d}z\,\partial_b {\cal W}(\phi_{\rm sol}) \, 
\frac{\partial\phi_{\rm sol}^b}{\partial m^j}
 \, \,
= - \frac{\partial}{\partial m^j}\!\int\!{\rm d}z\left[ \,{\cal
W}(\phi_{\rm sol})-{\cal W}(\phi_{\rm
sol})_{m=m_*}\right]\,,
\label{A2}
\end{equation}
where we use the fact that the soliton solution depends on
the spatial coordinate  only through 
the combination $z-z_0$, to replace
${\partial \phi_{\rm sol}^a}/{\partial m^0}$
by ${\partial \phi_{\rm sol}^a}/{\partial z}$,
which, in turn, can be replaced by $\, g^{ac}\,\partial_c \,{\cal
W}(\phi_{\rm sol})$ by virtue of Eq.~(\ref{AAone}).
We also regularized 
the integral on the right-hand side
of Eq.~(\ref{A2})  by subtracting from the integrand
the superpotential at some fixed values of the moduli 
$m = m_*$. 

Considering Eq.~(\ref{A2}) for $h_{00}$ we get 
\begin{equation}
h_{00}= - \frac{\partial}{\partial m^0}\!\int\!{\rm d}z\left[ \,{\cal
W}(\phi_{\rm sol})-{\cal W}(\phi_{\rm
sol})_{m=m_*}\right]\,.
\end{equation}
Having in mind $h_{00}={\cal Z}$ we define the modulus $m^0$ as  
\begin{equation}
m^0= -\frac{1}{{\cal Z}} \!\int\!{\rm d}z\left[ \,{\cal
W}(\phi_{\rm sol})-{\cal W}(\phi_{\rm
sol})_{m=m_*}\right]\,.
\label{mzero}
\end{equation}
With this definition it is clear that  
\begin{equation}
h_{0j} =\frac{\partial m_0}{\partial m^j}=0\,,\quad (j=1,...,n)\,.
\end{equation}

Thus, the Lagrangian 
describing the moduli dynamics has the form
\begin{equation}
L\left({ M}^{n +1}\,\right)=-{\cal Z}
+\frac{{\cal Z}}{2}\left[(\dot
z_0)^2 +i
\,\eta\,\dot\eta\,
\right] +
L\left({\cal M}^{n }\right)\,,
\label{Athree}
\end{equation}
where $L\left({\cal M}^{n }\right)$
is the Lagrangian of the internal moduli, both bosonic and fermionic,
a sigma-model quantum mechanics on ${\cal M}^{n }$. We see, that the motion of
the center of mass (together with its fermionic partner) is factored out, 
and we only need to consider the dynamics on ${\cal M}^{n}$.

The simplest case $n=0$ was already analyzed in Sec.\,5.1. In this case the
quantum moduli  dynamics is trivial, and the single state BPS multiplet does
exist, ${\rm Ind}_{\,\cal Z}\,(Q_2/Q_1)=1$.  
For $n\ge1$ one must differentiate between even and odd $n$.  

For odd $n$ the total number $n+1$ of the fermion moduli is even. Under
quantization these $n+1$ moduli become $\gamma $ matrices (multiplied by 
frames, as in Eq.~(\ref{modsigma})) satisfying 
the Clifford
algebra with an even number of generators. 
All $\gamma$'s are multiplied by frames, as in Eq. (93)
Taking the product of all these $\gamma $ matrices we get the matrix $\gamma^{n+2}=\prod \gamma^i$ which anticommutes with all $\gamma^i$, ($i=1,\ldots,n+1$).
This $\gamma^{n+2}$,  an analog of $\gamma^5$ 
in four dimensions,  represents $(-1)^F$, i.e., all multiplets are long.

Consider now  a less trivial case of even $n$ ---
only in this category can one  expect to
find Tr$\,Q_1\neq 0$.
Quantization of $L\left({\cal M}^{n}\right)$
is standard. All operators
act in 
the Hilbert space of the spinor wave functions $\Psi_\alpha (m)$, 
where $\alpha=1,\ldots, 2^{n/2}$. The operators $m^i$ act as multiplication,
while $\dot m_i$ become matrix-differential operators.
The fermion moduli of the reduced moduli space 
become $\gamma$ matrices of dimension
$2^{n/2}\times 2^{n/2}$. 
 The matrix $\gamma^{n+1}=\prod_{i=1}^{i=n} \gamma^i$ 
is used to represent the remaining fermion modulus, a partner of translation. 
On the moduli space ${\cal M}^{n}$ 
the supercharges (\ref{supercur}) take the form
\begin{equation}
Q_1=\sqrt{2{\cal Z}}\, \gamma^{n+1}\,,\qquad
 Q_2=-\frac{i}{\sqrt{2}}\;\gamma^j\,\nabla_j \,,
\label{qq22}
\end{equation}
where the covariant derivative $\nabla_j$ includes spin connection.
The expression for $Q_2$ is in fact the Dirac operator $i\!\!\not \!
\nabla$ on
${\cal M}^n$. Moreover, the Hamiltonian takes the form,
\begin{equation}
H-{\cal Z}=Q_2^2=\frac{1}{2}\,(i\!\!\not \!\nabla)^2=
-\frac{1}{2}\,\nabla^j\nabla_j+
\frac{1}{8}\,\tilde R\,,
\label{lich}
\end{equation}
 where $\tilde R$ is the 
curvature in the soliton moduli space, and we again  used the 
Lichnerowicz formula (cf. Eq.~(\ref{hamilt})).

From Eqs.~(\ref{qq22}), (\ref{lich}) it is clear that the BPS soliton states 
are in correspondence with the zero modes of the Dirac operator
$i\!\!\not \! \nabla$ on
${\cal M}^{n }$.
 The index ${\rm Ind}_{\,\cal
Z}\,(Q_2/Q_1)$ we defined in Eq.~(\ref{Ind}) becomes the square of the index 
of the
Dirac operator
\begin{eqnarray}
&&{\rm Ind}_{\,\cal Z}\,(Q_2/Q_1)=\left\{{\rm Ind}\,(i\!\!\not \!\nabla)_{{\cal
M}^{n}}\right\}^2, \nonumber\\[2mm]
&&
{\rm Ind}\,(i\!\!\not \! \nabla)_{{\cal M}^{n}}={\rm Tr}
\left[\,\gamma^{n+1}
\exp\left(\beta\!\!\not \! \nabla^2
\right)\right]_{{\cal M}^{n}}.
\label{IndDir}
\end{eqnarray}

Equation (\ref{lich}) shows that if the curvature $\tilde R$ is positive 
everywhere 
on the soliton moduli space the Dirac operator has no zero modes, its index
vanishes, and so does the index ${\rm Ind}_{\,\cal Z}\,(Q_2/Q_1)$. Thus, 
there are no BPS solitons in this case. An explicit example  
is provided by a sigma model on 
$S^3$. 

Moreover, the situation turns out to be general in the hybrid models: the geometry of the reduced moduli space is similar to spherical, the index of the Dirac operator vanishes for any compact ${\cal M}^{n}$ with $n\ge 1$. 
The proof due to P.~Pushkar' is presented  in 
Appendix. 

A comment is in order here concerning the vanishing of the index of
the Dirac operator on even-dimensional reduced moduli spaces
${\cal M}^{2\ell}$ (with integer $\ell$). Naively one might
be tempted to think that since all $\gamma$ matrices are traceless the 
index in Eq.~(\ref{IndDir})
vanishes automatically, irrespective of the properties
of ${\cal M}^{2\ell}$. It is well known that this naive conclusion is
wrong --- a more careful consideration is necessary.

The moduli dynamics is governed by the Lagrangian
$$
L({\cal M}^{2\ell})=\frac{1}{2}\,h_{ij}(m)\left( \dot{m}^i \dot{m}^j +i\eta^i {\cal D}_t \eta^j\right)
$$
where the metric $h_{ij}$ is that for the reduced moduli space given in
Eq.~(\ref{Aone}) with $i,j = 1,2,...,2\ell$.
As we have already mentioned, after
quantization the wave functions become
spinors $\Psi_{\alpha}(m)$, while $\eta$'s turn out to be $\gamma$
matrices
(times the frames).

The index of the Dirac operator is the
regularized trace of $\gamma^{2\ell+1}$ over the space of
wave functions, see Eq.~(\ref{IndDir}). Naively, one might think that 
the space of the wave functions is a product
of the spinor representation of the $2\ell$-dimensional Clifford algebra
and
the space ${\cal C}$ of smooth functions on ${\cal M}^{2\ell}$.
If so, the trace of $\gamma^{2\ell+1}$ over the space of the wave
functions would be automatically zero.

In fact, the space of wave functions is a product of the two spaces
mentioned above
only locally!  The manifold ${\cal M}^{2\ell}$ should be thought of as
covered
by open sets, and when we go from one open set to another,
generally speaking  we need  to rotate the spinor representation with
the
${\rm Spin}(2\ell)$-valued function of $m^i$. This means that the wave
functions
$\Psi_{\alpha}(m)$ are sections of the spinor bundle that
is generically nontrivial. The index of the Dirac operator is one of
the characteristics
that reflects this nontriviality. It might be nonzero would
${\cal M}^{2\ell}$ be similar to $CP_2$ or $K3$.
The central point of the Pushkar' theorem outlined in Appendix
is that the geometry of the soliton reduced moduli
space is similar to  spherical and cannot be similar to that of  
$CP_2$ or $K3$.

Thus, for $n\ge 1$ the soliton multiplets 
are long and generically non-BPS, $M>{\cal Z}$. If, for 
accidental or other reasons, they are still BPS saturated,  they form a 
reducible representation. For example, in \ntwo models the index  
${\rm Ind}_{\,\cal Z}\,(Q_2/Q_1)$ vanishes while the BPS states do exist. 
From the standpoint of \none
they form a reducible representation for which $(-1)^F$ is well defined.

Our  consideration refers to the case of compact ${\cal M}^{n}$. 
Generally speaking, ${\cal M}^{n}$ may be noncompact.
Noncompact geometry of ${\cal M}^{n}$  may emerge, for example,  
if there is an 
extra critical point $C$ such that
${\cal Z}_{AB}={\cal Z}_{AC}+{\cal Z}_{CB}$.
Physically it means that there is an infinite degeneracy of the quantized 
soliton states.

\section{Conclusions}
\label{sec:conlusion}

We analyzed a wide class of hybrid  models with \none supersymmetry and central
charges in (1+1) dimensions. 
For the BPS states only one out of two supercharges is realized
nontrivially which leads to one-dimensional irreducible representations
of the
superalgebra. The non-BPS supermultiplets are two-dimensional.
Our main topic was the soliton multiplet structure in various models at weak
coupling (quasiclassical approach).  

We introduced and thoroughly discussed
the index Tr$\,Q_1/\sqrt{2{\cal Z}}$ which counts the supershort 
(single-state) supermultiplets.
It was demonstrated that nonvanishing Tr$\,Q_1$ implies the loss
of the fermion parity $(-1)^F$. We showed that Tr$\,Q_1$ is related to 
the index of the Dirac operator on the reduced moduli space. The geometry 
of this space is similar to spherical. It implies that the index vanishes except the very special case when the reduced moduli space is a point.

The vanishing index implies long multiplets which may or may not be BPS 
saturated. It is clear that the BPS saturation is not protected for long 
multiplets. We demonstrated that indeed  quantum corrections destroy BPS saturation in many cases, by  calculating the 
supercharge $Q_2$ at
one loop. Classically vanishing $Q_2$ becomes nonzero 
at one loop.   This leads to $M-{\cal Z}\neq 0$ at two loops. 
 In special cases,
where $Q_2$ remains zero in perturbation theory, 
$Q_2 \neq 0$  may be generated nonperturbatively.\cite{Tonnis}

What lessons have we learned from the study of the \none  theories?
The main lesson is that of the fermion quantization in the case when the 
number of fermion zero modes is odd. Let us remind that 
the only consistent approach
to  fermions in field theory is based on Berezin's 
holomorphic quantization which implies the number of the fermion 
degrees of freedom is even. 

How can it be consistent with the fact that 
the BPS-saturated irreducible representations of
${\cal N}=1$ centrally extended superalgebra
are one-dimensional?  The resolution is as follows:
if the theory is explicitly regularized both
in the ultraviolet and infrared it 
 becomes a quantum mechanics
of a large number of variables; the number of the fermion
variables is necessarily even,
the supermultiplets are reducible 
and $(-1)^F$ is preserved. Thus, in fully regularized theory the number 
of the fermion zero modes is always even, and so is
the number of states in the supermultiplet. Moreover, the 
BPS saturation is not protected so we can say that  
there are no BPS states in the fully regularized theory.

We gave  clear-cut illustrations to this point.
In Sec.~\ref{sub:circle} the spatial dimension was compactified
onto a circle. The soliton had two zero modes -- one 
localized on the soliton, another on the other side of the circle.
Another example is given by the softly broken \ntwo theory, discussed in  
Sec.~\ref{sec:2sf}. 
 There  we could see 
how the second zero mode becomes delocalized
once the ${\cal N}=2$ breaking
parameter becomes large enough
(more exactly the localization
of the second zero mode shifts 
away from the soliton,
to the edge of the box).

In spite of the ``evenness'' of the total
number of the 
fermion  modes, in the limit $L\to\infty$
it may (and actually does) happen
that in the physical subsector of the Hilbert space
there remains an odd number of the fermion moduli.
If some of the modes are localized at the boundaries 
of the large box, the corresponding states are 
unobservable in any physical local measurement.
What is observable are the localized states
associated with the soliton. If the number of
fermion modes localized on the soliton
is odd, we arrive to an abnormal situation.
An explicit example was given in Sec. 5.
In this case we get the multiplet
shortening, and the BPS saturation is implemented.
The number of such short multiplets in the physical subsector
is counted by Tr$\,Q_1/\sqrt{2{\cal Z}}$.

Our results 
naturally ``blend in" into a general picture.  The BPS saturation was
studied in detail in the \ntwo ~theories in (1+1) dimensions\,\cite{Ntwo}
and in
the \ntwo\!\!,
\nfour ~theories in (3+1)-dimensions.  
The assertion that under the full regularization (UV and IR) there are no 
short multiplets is general, it is applicable to higher dimensions and 
higher supersymmetries. Extra states are localized away from the soliton 
center. What is specific for \none in two dimensions, where the number of 
supercharges is minimal, is the odd number of the soliton fermion moduli 
(in higher dimension it is always even). This may lead to the loss of
$(-1)^F$ in the physical sector.

In a broader context, we found another example of a remarkable phenomenon
first discussed by Witten\,\cite{Ed} -- supersymmetry without the 
full fermion-boson
degeneracy. 
 If such theories could be found in four dimensions,
this would be ``a dream came true."  

Witten's example is 
 in the context of 2+1 supergravity with the conic geometry.
Out of four supercharges of the model two
supercharges annihilate the BPS solitons. The other two supercharges
produce the fermion zero modes. Without gravity these modes are normalizable
which leads to two-component short supermultiplet of \ntwo. With gravity switched on the fermion
modes become non-normalizable, implying the single-state supermultiplet.
This means that in the physical sector of the localized states all 
supercharges 
act on the soliton trivially.

In our \none examples of the single-state supermultiplet one of two 
supercharges is realized nontrivially, $Q_1=\pm\sqrt{2{\cal Z}}$.  In terms 
of modes there is one
normalizable fermion  mode.  In Witten's case all fermion zero modes
run away to the boundary, while in our case 
one mode  is localized on the soliton, and 
the other at the boundary.
Similar
run-away behavior of the modes occurs in the phenomenon of the
fractional charge and other phenomena known in solid state physics.

\vspace{-3mm}
\section*{Acknowledgments}
\addcontentsline{toc}{section}{\numberline{}Acknowledgments}

Our study was stimulated by a comment made by Ken Intriligator who
noted, in
response to Ref.\,\protect\citebk{1}, that there was no doubling of short
multiplets in
\ntwo ~in 1+1. We are grateful to him for this remark. We would like to
thank 
 Paul Fendley, Leonid Glazman, Fred Goldhaber, Anatoly Larkin,
Bob Laughlin, Peter van Nieuwenhuizen, Pyotr Pushkar',  Adam Ritz, Andrei
Smilga,
 Boris Shklovsky, Alexei Tsvelik,  Misha Voloshin, Edward Witten, and Alexander
Zamolodchikov for  valuable discussions.

The work of A.L. was supported in part by RFFI grant 00-02-16530,
Support for Scientific Schools grant 00-15-96-557,  and INTAS grant
99-590, the work of M.S. and A.V. was supported in part by  DOE  grant 
DE-FG02-94ER408.

\vspace*{-3mm}
\section*{Appendix (by P.~Pushkar'): Vanishing of the index of the Dirac operator on compact reduced moduli space}
\label{sec:append}
\addcontentsline{toc}{section}{\numberline{}Appendix (by P.~Pushkar'): Vanishing of the index of the Dirac operator on compact reduced moduli space}

The proof presented in this Appendix belongs to  P.~Pushkar'.

The index of the Dirac operator on the reduced moduli space ${\cal M}^{n}$ 
is known to be equal to
$\hat{A}$-genus and can be expressed as an
integral of polynomial of Pontriagin classes
along ${\cal M}^{n}$.
In order to show that the index vanishes we will show (in {\em Statement 3})
 that tangent bundle to
the space ${\cal M}^{n}$ of nonparametrized trajectories
is such, that its sum with the trivial bundle is a trivial bundle.
Really, then its Pontriagin classes (of nonzero degree) should vanish.

Here we assume that there are no other critical values between values
of the initial and final critical points of superpotential, thus,
the space ${\cal M}^{n}$ is compact.
We will also assume that dimension of ${\cal M}^{n}$ is nonzero.
Let us take the equi(super)potential surface $L$ defined by ${\cal W}=c$ 
where the constant  $c$ is between the values of ${\cal W}$
at the initial and final points.
The gradient trajectories coming out of the initial point intersect the 
surface $L$
and produce a sphere $S^{\rm initial}$. The antigradient trajectories coming out
of the final point also intersect $L$ and produce another sphere 
$S^{\rm final}$.
The space ${\cal M}^{n}$ is an intersection of these two spheres.
\\[2mm]
{\em Statement 1:} The normal bundles of $S^{\rm initial}$ and 
$S^{\rm final}$  in $L$ are trivial.
\\[2mm]
For instance, to show this for $S^{\rm initial}$
let us move $L$ close to the initial critical
point (it would be a homotopy that should not change the triviality
of the bundle).  In the vicinity of the critical point we can
replace the function by its quadratic approximation -- then the
triviality of the normal bundle becomes obvious.
\\[2mm]
{\em Statement 2:} The normal bundle to ${\cal M}^{n}$ in $S^{\rm initial}$ 
is trivial.
\\[2mm]
Indeed, this bundle is a restriction
(to ${\cal M}^{n}$)  of the normal bundle of $S^{\rm final}$
in $L$, and the latter is trivial due to {\em Statement 1}.
Since the restriction of the trivial bundle is trivial, we
have proved {\em Statement 2}.
\\[2mm]
{\em Statement 3:} The tangent bundle to ${\cal M}^{n}$ plus the trivial bundle
is a trivial bundle.
\\[2mm]
Suppose that ${\cal M}^{n}$ is different from the total sphere 
$S^{\rm initial}$,
then it is a submanifold of the Euclidean space.
The tangent bundle to ${\cal M}^{n}$ plus the normal bundle
(that is trivial due to {\em Statement 2}) gives a restriction
on ${\cal M}^{n}$ of the tangent
bundle to the Euclidean space (that is obviously trivial).
Since the restriction of the trivial bundle is trivial we have proved
{\em Statement 3} for ${\cal M}^{n} \neq S^{\rm initial}$.
If ${\cal M}^{n}$ is a sphere $S^{\rm initial}$, then it could be obviously embedded in
the Euclidean space with the trivial normal bundle. This completes the proof
of {\em Statement 3}.

The generation function for Pontriagin classes for a trivial bundle
is a class of degree zero.
The generation function for Pontriagin classes of the sum of
bundles is the product of the generation functions of each bundle.
Thus, the only nonvanishing Pontriagin class on ${\cal M}^{n}$ is of degree
zero,  and the integral of the polynomial of Pontriagin classes
along ${\cal M}^{n}$ will give zero.

\section*{References}
\addcontentsline{toc}{section}{\numberline{}References}

\end{document}